\documentclass[longauth]{aa} 

\newcommand{\microm}{$\mu$m}

\newcommand{\uchii}{{UC\ion{H}{ii}~}}
\newcommand{\hii}{\ion{H}{ii}~}
\newcommand{\herschel}{{\it Herschel}}

\usepackage{color}

\usepackage{natbib}
\usepackage{amsmath} 
\usepackage{subfig} 

\usepackage{graphicx}
\usepackage{rotating}
\usepackage{txfonts}
\usepackage[normalem]{ulem}
\begin{document}

\titlerunning{Mon~R2 \hii regions structure and age from \herschel}
\title{
From forced collapse to \hii region expansion in Mon~R2: \\ Envelope density structure and age determination with \herschel 
\thanks{\emph{Herschel}\ is an ESA space observatory with science instruments provided by European-led Principal Investigator consortia and with important participation from NASA}
   }

   \author{
          P. Didelon\inst{1}
          \and
          F. Motte\inst{1}   \and
          P.~Tremblin\inst{1,2,3}   \and
          T. Hill\inst{1,4}        \and
          S.~Hony\inst{1,5}  \and
          M.~Hennemann\inst{1}  \and
          P.~Hennebelle\inst{1}  \and
          L. D.~Anderson\inst{6}   \and
          F.~Galliano\inst{1}   \and
          N.~Schneider\inst{1,7}   \and
          T.~Rayner\inst{8}   \and          
           K.~Rygl\inst{9}   \and
          F.~Louvet\inst{1,10}        \and
           A.~Zavagno\inst{11}             \and
          V.~K\"onyves\inst{1}    \and
           M.~Sauvage\inst{1,}   \and
          Ph.~Andr\'e\inst{1}  \and
          S.~Bontemps\inst{7}  \and
           N.~Peretto\inst{1,8}  \and
          M.~Griffin\inst{8,12} \and
          M.~Gonz\'alez\inst{1}  \and
          V.~Lebouteiller\inst{1}  \and
          D. Arzoumanian\inst{1}  \and           
          J.-P.~Bernard\inst{13} \and
          M.~Benedettini\inst{14}  \and
          J.~Di~Francesco\inst{15, 16} \and
           A.~Men'shchikov\inst{1} \and
           V~. Minier\inst{1}            \and
           Q.~Nguy$\tilde{\hat{\rm e}}$n~Lu{\hskip-0.65mm\small'{}\hskip-0.5mm}o{\hskip-0.65mm\small'{}\hskip-0.5mm}ng\inst{1,17}   \and   
           P.~Palmeirim\inst{1}  \and
           S.~Pezzuto\inst{14}  \and
           A.~Rivera-Ingraham\inst{18}  \and
           D.~Russeil\inst{11}     \and
          D.~Ward-Thompson\inst{8,19} \and
          G. J.~White\inst{20}   
         }

   \institute{Laboratoire AIM, CEA/IRFU CNRS/INSU Universit\'e Paris Diderot, CEA-Saclay, 91191 Gif-sur-Yvette Cedex, France\\
              \email{pierre.didelon@cea.fr}
                \and 
        Astrophysics Group, University of Exeter, EX4 4QL Exeter, UK
                \and 
        Maison de la Simulation, CEA-CNRS-INRIA-UPS-UVSQ, USR 3441, Centre d\'\ \'etude de Saclay, 91191 Gif-Sur-Yvette, France 
              \and 
        Joint ALMA Observatory, 3107 Alonso de Cordova, Vitacura, Santiago, Chile
             \and 
        Universit\"{a}t Heidelberg, Zentrum f\"{u}r Astronomie, Institut f\"{u}r Theoretische Astrophysik, Albert-Ueberle-Str. 2, 69120 Heidelberg, Germany
              \and 
              Department of Physics and Astronomy, West Virginia University, Morgantown, WV 26506, USA ; 
             Also Adjunct Astronomer at the National Radio Astronomy Observatory, P.O. Box 2, Green Bank, WV 24944, USA
             \and 
             Universit\'e de Bordeaux, OASU, Bordeaux, France             
            \and 
             Cardiff University, Wales, UK  
            \and  
            European Space Research and Technology Centre (ESA-ESTEC), Keplerlaan 1, PO Box 299, NL-2200 AG Noordwijk, the Netherlands
            \and 
            Cerro Calan, Observatorio Astronmico Nacional, Camino el Observatorio, 1515, Las Condes, Chile
             \and 
             Laboratoire d'Astrophysique de Marseille, CNRS/INSU--Universit\'e de Provence, 13388 
             Marseille cedex 13, France
             \and 
             Queen Mary + Westf. College, Dept. of Physics, London, UK
             \and 
             CESR, Toulouse, France             
            \and  
            INAF-Istituto di Astrofisica e Planetologia Spaziali, via Fosso del Cavaliere 100, I-00133 Rome, Italy
            \and 
            National Research Council of Canada, Herzberg Institute of Astrophysics, 5071 West Saanich Rd., Victoria, BC, V9E 2E7, Canada
            \and  
            University of Victoria, Department of Physics and Astronomy, PO Box 3055, STN CSC, Victoria, BC, V8W 3P6, Canada
            \and 
            National Astronomical Observatory of Japan, Chile Observatory, 2-21-1 Osawa, Mitaka, Tokyo 181-8588, Japan
            \and  
            Universit\'e de Toulouse, UPS-OMP, IRAP, F-31028 Toulouse cedex 4, France ; CNRS, IRAP, 9 Avenue colonel Roche, BP 44346, F-31028 Toulouse cedex 4, France
           \and 
             Jeremiah Horrocks Institute, University of Central Lancashire,
             Preston, Lancashire, United Kingdom
            \and  
            Department of Physical sciences, The Open University, Milton Keynes, UK ; RALspace, The Rutherford Appleton Laboratory, Chilton, Didcot, UK
             }


\abstract
{The surroundings of \hii regions 
can have a profound influence on their development, morphology, and evolution.
This paper explores the effect of the environment on \hii regions in the MonR2 molecular cloud.
}   
{We aim to investigate the density structure of envelopes surrounding \hii regions and to determine their collapse and {ionisation} expansion ages. 
The Mon~R2 molecular cloud is an ideal target since it hosts an \hii region association, which has been imaged by the \emph{Herschel} PACS and SPIRE cameras as part of the HOBYS key programme.
}
{Column density and temperature images derived from \emph{Herschel} data were used together to model the structure of \hii bubbles and their surrounding envelopes. 
The resulting observational constraints were used to follow the development of the Mon~R2 ionised regions with analytical calculations and numerical simulations.}
{The four hot bubbles associated with \hii regions 
are surrounded by dense, cold, and neutral gas envelopes, which are partly embedded in filaments. 
The envelope's radial density profiles are reminiscent of those of low-mass protostellar envelopes. 
The inner parts of envelopes of all four \hii regions could be free-falling because they display {shallow density profiles: $\rho(r)\propto r^{-q}$ with $q \leqslant 1.5$}. 
As for their outer parts, the two compact \hii regions show a $\rho(r)\propto  r^{-2}$ profile, which is typical of the equilibrium structure of a singular isothermal sphere. 
In contrast, the central UC\hii\ region shows a steeper outer profile, $\rho(r)\propto r^{-2.5}$, that could be interpreted as material being forced to collapse,  
where an external agent overwhelms the internal pressure support.}
{The size of the heated bubbles, the spectral type of the irradiating stars, and the mean initial neutral gas density are used to estimate the ionisation expansion time,
$t_\text{exp} \sim 0.1$~Myr, for the {dense} \uchii and compact \hii regions and $\sim0.35$~Myr  for the extended \hii region. Numerical simulations with and without gravity show that the so-called lifetime problem of \hii regions is an artefact of theories that do not take their surrounding neutral envelopes with slowly decreasing density profiles into account. 
The envelope transition radii between the {shallow}  
and steeper density profiles are used to estimate the time elapsed since the formation of the first protostellar embryo, 
$t_\text{inf}\sim 1$~Myr, for the ultra-compact, $1.5-3$~Myr for the compact, and greater than $\sim$6~Myr for the extended \hii regions. 
These results suggest that the time needed to form a OB-star embryo and to start ionising the cloud, plus the quenching time due to the large gravitational potential amplified by further in-falling material, dominates the ionisation expansion time by a large factor. 
Accurate determination of the quenching time of \hii regions would require additional small-scale observationnal constraints and numerical simulations including 3D geometry effects.
} 


   \maketitle
%
\section{Introduction}\label{sec:intro}

Molecular cloud complexes forming high-mass stars are heated  and structured by newly born massive stars and nearby OB clusters. 
This is especially noticeable in recent  \emph{Herschel} 
observations obtained as part of the HOBYS key programme, which is dedicated to high-mass star formation \citep[][see http://hobys-herschel.cea.fr]{motte2010,motte2012}. 
The high-resolution (from 6\arcsec\ to 36\arcsec), high-sensitivity observations obtained by \emph{Herschel} provide, through spectral energy distribution (SED) fitting, access to temperature and column density maps covering entire molecular cloud complexes. 
It has allowed us to make a detailed and quantitative link between the spatial and thermal structure of molecular clouds for the first time.
Heating effects of OB-type star clusters are now clearly observed to develop over tens of parsecs and up to the high densities (i.e. $\sim$$10^5$~cm$^{-3}$) of starless cores 
\citep[][]{Schneider2010SI,Hill2012M16, Roccatagliata2013}. 
Long known at low- to intermediate-densities, it has recently been reported  
that cloud compression and shaping lead to the creation of very high-density filaments 
hosting massive dense cores \cite[$10^5-10^6$~cm$^{-3}$,][]{minier2013, Tremblin2013}. 
While triggered star formation is obvious around isolated \hii regions \citep{zavagno2010,anderson2012,Deharveng2012}, it has not yet been unambiguously established 
as operating across whole cloud complexes 
\citep[see][]{Schneider2010SI,Hill2012M16}.

The Mon~R2 complex hosts a group of B-type stars, {three of them} associated with a reflection nebula \citep{vdb1966}. 
Located 830~pc from the Sun,  it spreads over 2~deg ($\sim$30~pc at this distance) along an east-west axis and has an age of $\sim$$1-6 \times 10^6$~yr (\citealt{HR76}, see Plate V). 
This reflection nebulae association corresponds to the Mon~R2 spur seen in CO \citep{wilson2005}. 
An infrared cluster with  a similar age \citep[$\sim$$1-3 \times 10^6$~yr,][]{aspen1990,carpenter1997} covers the full molecular complex 
and its \hii region association area \citep[Fig.~1 in][]{Gutermuth2011}. 
The western part of the association hosts the most prominent object in most tracers. This UC \hii region \citep{Fuente2010}, which is 
powered  by a B0-type star \citep{downes75}, is associated with the infrared source Mon~R2 IRS1 \citep{massi85, henning1992}. 
The \uchii\ region drives a large bipolar outflow \citep{Richardson1988,Meyers-Rice_Lada1991,xie1994} that is oriented NW-SE and is approximately aligned with the rotation axis of the full cloud found by  \cite{loren1977}.
The molecular cloud is part of a CO shell with $\sim$26~pc size whose border encompasses the \uchii region \citep{wilson2005}. 
The shell centre is situated close to  VdB72/NGC2182  \citep{xie1994,wilson2005} at the border of the temperature and column density map area defined by  the Herschel SPIRE and PACS common field of view (see e.g. Fig.~\ref{fig:id_cdens-extraArea4prof}). 
\citet*{loren1977} also observed CO motions that he interpreted as tracing the global collapse of the molecular cloud with an infall speed of a few km\,s$^{-1}$. 
This typical line profile of infall has also been seen locally in CO and  marginally in $^{13}$CO 
near the central UC\hii\ region thanks to higher resolution observations \citep{Tang2013}. 
Younger objects, such as molecular clumps seen in CO, H$_2$CO, and HCN, for example, have been observed in this molecular cloud \citep[e.g.][]{Giannakopoulou}.

In this paper, we constrain the evolutionary stages of the Mon~R2 \hii regions through their impact on 
the temperature and the evolution of the density structure of their surrounding 
neutral gas envelopes. 
The evolution and growth of the H~{\sc ii} region as a 
function of age depends strongly on the density structure of the 
surrounding environment. 
If a simple expansion at the thermal sound speed determines their sizes, the number of UCH~{\sc ii} regions observed in the galaxy exceeds expectations.
This is the so-called lifetime problem \citep{Wood1989,Churchwell2002}
{
that partly arises from a mean density and a mean ionising flux representative of a sample but perhaps not adapted to an individual object}.
To precisely assess the UCH~{\sc ii} lifetime problem, we need an accurate measure of the 
observed density profile of the neutral envelope outside of the ionised gas

One goal of this paper is to use the sensitive \emph{Herschel} 
FIR (Far InfraRed) and sub-mm photometry of the compact H~{\sc ii} regions in MonR2 to 
constrain these profiles and subsequently estimate the H~{\sc ii} regions ages based on {\it \emph{measured}} profiles 
{
and the corresponding average density}. 
Moreover, \emph{Herschel} measurements are sensitive enough that we can even investigate significant changes in the density profile
with radial distance. Such changes may be signposts of dynamical processes, such as infall or external compression. 

The paper is organised as follows. 
\emph{Herschel} data, associated column density, dust temperature images, and additional data are presented in Sect.~\ref{s:obs}. 
Section~\ref{s:hii_and_shell} gives the basic properties of \hii regions and the density structure of their surrounding envelopes. 
These constraints are used  to estimate  \hii region expansion in Sect.~\ref{sec:ionisationExpansion} and 
the age of the protostellar accretion in Sect.~\ref{sec:HIIRage},
to finally get a complete view of the formation history of these B-type stars.

\begin{figure}[htbp]
 \begin{minipage}{1\linewidth}
\includegraphics[height=1.\hsize]{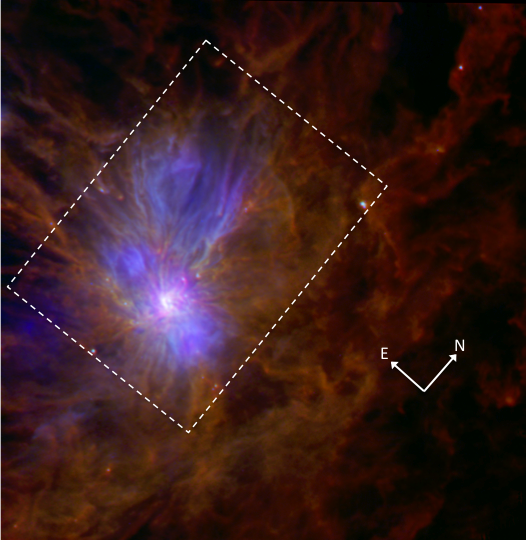} 
 \end{minipage}   
\hfill 
\caption{Three-colour  \emph{Herschel} image of the Mon~R2 molecular complex using  70\,\microm\,(blue), 160\,\microm\,(green), and 250\,\microm\,(red) maps. The shortest wavelength (blue) reveals the hot dust associated with \hii regions and protostars. The longest wavelength (red) shows the cold, dense cloud structures, displaying a filament network. 
{The dashed rectangle locates the four \hii region area shown in Fig.~\ref{fig:hotRcoldTmaps}}
 \label{fig:3col}}
\vspace{-.4cm}
\end{figure}
 

\section{Data}  
\label{s:obs}

\subsection{Observations and data reduction}
Mon~R2 was observed  by \emph{Herschel} \citep{Pilbratt2010} 
on September 4, 2010 (OBSIDs: 1342204052/3), 
as part of the HOBYS key programme \citep{motte2010}. 
The parallel-scan mode was used with the slow scan-speed (20\arcsec/s), allowing simultaneous observations with the PACS \citep[70 and 160\,\microm;][]{Poglitsch2010} and SPIRE \citep[250, 350, 500\,\microm;][]{Griffin2010} instruments at five bands. 
To diminish scanning artefacts, two nearly perpendicular coverages of $1.1^\circ \times 1.1^\circ$ were obtained. The data were reduced with version 8.0.2280 of HIPE\footnote{
HIPE is a joint development software by the Herschel Science Ground Segment Consortium, consisting of ESA, the NASA \emph{Herschel} Science Center, and the HIFI, PACS, and SPIRE consortia.}
standard steps of the default pipeline up to level 1, including calibration and deglitching, and were applied to all data. The pipeline was applied (level-2) further, including destriping and map making, for SPIRE data. 
To produce level-2 PACS images, we used the {\it Scanamorphos} software package\footnote{
http://www2.iap.fr/users/roussel/herschel/index.html} v13, which performs baseline and drift removal before regriding \citep{Roussel2012}. 
The resulting maps are shown in Rayner et al. (in prep.). 
The map angular resolutions are $6\arcsec-36\arcsec$, which correspond to $0.025-0.15$~pc at the distance of Mon~R2.

The \emph{Herschel} 3-colour image of the Mon~R2 molecular complex shows that the central UC\hii region,  
seen as a white spot in Fig.~\ref{fig:3col}, dominates and irradiates its surroundings. 
Three other \hii regions exhibit similar but less pronounced irradiating effects (see the bluish spots in Fig.~\ref{fig:3col}). 
These \hii regions develop into a clearly structured environment characterised by cloud filaments.

\begin{figure*}[htbp]
 \includegraphics[height=.7\hsize]{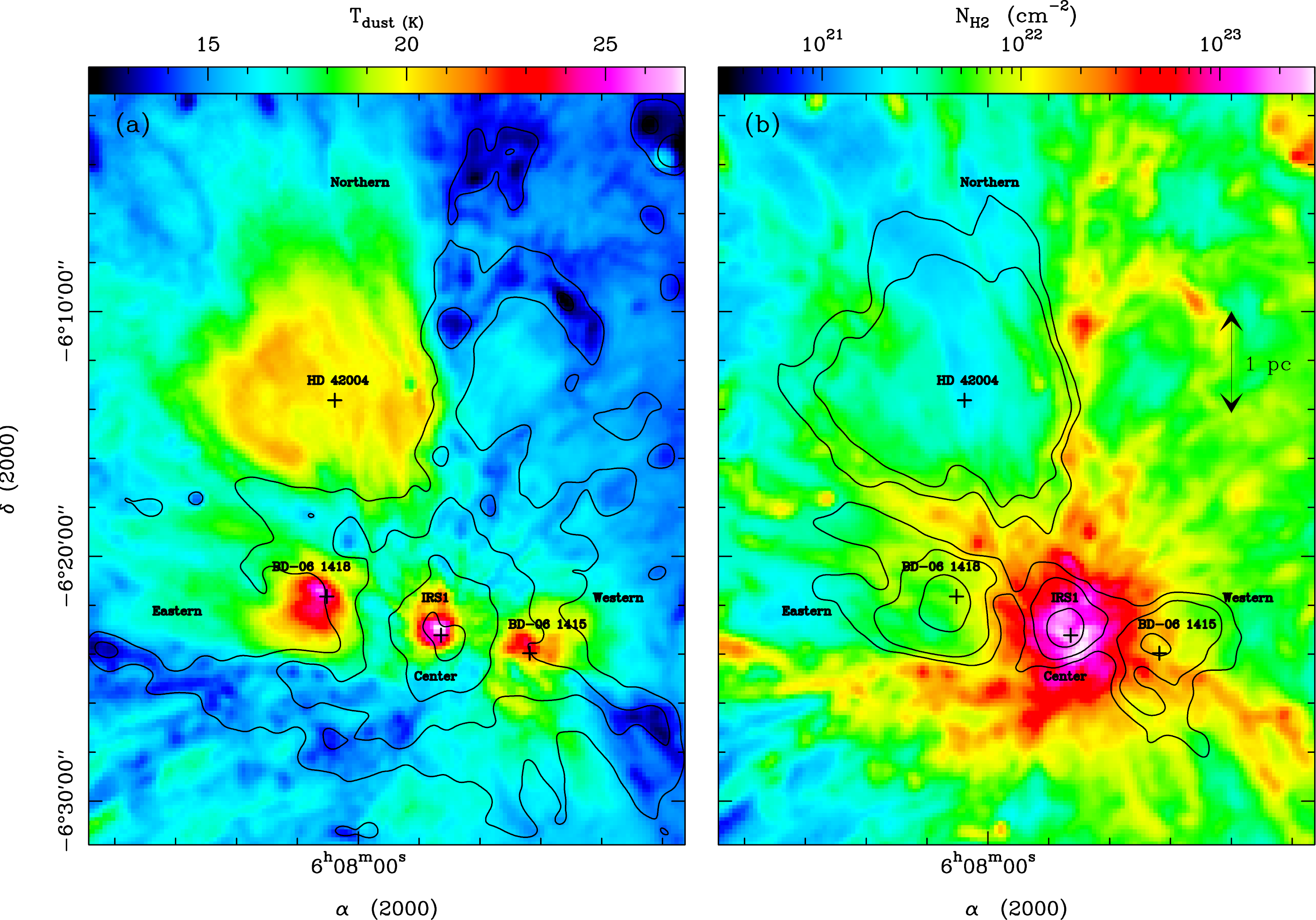} 
\hfill 
\vspace{-.6cm}
\caption{
Dust temperature  (colours in {\bf a} and contours in {\bf b}) and column density (colours in {\bf b} and contours in {\bf a}) maps 
of the central part of the Mon~R2 molecular cloud complex, covering the  
central, western, eastern, and northern \hii regions with $36\arcsec$ resolution.
The heating sources of \hii regions are indicated {by a cross and with} their name  
taken from  \citet{racine68} and \citet{downes75}. 
Smoothed temperature and column density contours are 17.5, 18.5, and 21.5~K and 
$7\times10^{21}$, $1.5\times10^{22}$, $5\times10^{22}$, and $1.8\times10^{23}$~cm$^{-2}$.
\label{fig:hotRcoldTmaps}}
\vspace{-.5cm}
\end{figure*}

\subsection{Column density and dust temperature maps}\label{sec:colden}

The columm density and dust temperature maps of Mon~R2 were drawn using pixel-by-pixel SED fitting to a modified blackbody function with a 
single dust temperature \citep[][]{Hill2009}. 
We used a dust opacity law similar to that of \cite{Hildebrand1983}, assuming a dust spectral index of $\beta$\,=\,2 and a gas-to-dust ratio of 100, 
so that dust opacity per unit mass column density is given by $\kappa_{\nu}=0.1~(\nu/1000~{\rm GHz})^{\beta}$ cm$^2$\,g$^{-1}$, 
as already used in most HOBYS studies \citep[e.g.][]{motte2010,quang2013}. 
The 70 $\mu$m emission very likely traces small grains in hot PDRs and so was excluded from the fits 
because it does not trace the cold dust used to measure the gas column density \citep{Hill2012M16}.

We constructed both the {four-band} ($T_{\rm 4B}$) and {three-band} ($T_{\rm 3B}$) temperature images and associated column density maps 
by either using the four reddest bands (160 to 500$\mu$m)  
or only the 160, 250, and 350\,\microm\ bands of \emph{Herschel} \citep[cf.][]{hill2011, Hill2012M16}. 
Prior to fitting, the data were convolved to the  $36\arcsec$ or $25\arcsec$ resolution of the 500\,\microm~(resp. 350\,\microm) band and the zero offsets, obtained from comparison with Planck and IRAS \citep{bernard2010}, were applied to the individual bands. 
The quality of a SED fit was assessed using $\chi^2$ minimisation. Given the high quality of the \emph{Herschel} data, a  
reliable fit can be done even when dropping the 500\,\microm\ data point. 
For most of the mapped points, which have medium column density and average dust temperature values, $N_{\rm H2}$ and $T_{\rm dust}$ values 
are only decreased by 10\% and increased 5\%, respectively, {between $T_{\rm 4B}$  and  $T_{\rm 3B}$}.
In the following, we used the column density or temperature maps built from the four longest \emph{Herschel} wavelengths, except when explicitly stated. 
The high-resolution $T_{\rm 3B}$, $I_{\rm 70\mu m}$, and $I_{\rm 160\mu m}$ maps have only been used for size measurements, not for profile-{slope determinations}.

The resulting column density map shows that the central UC\hii region has a structure that dominates the whole cloud with column densities up to $N_{\rm H2} \sim2\times10^{23}$~cm$^{-2}$ (see Figs.~\ref{fig:hotRcoldTmaps}b and \ref{fig:id_cdens-extraArea4prof}). 
In its immediate surroundings, i.e. $\sim$1~pc east and west, as well as $\sim$2.5~pc north, the temperature map of Fig.~\ref{fig:hotRcoldTmaps}a shows two less dense \hii regions and one extended \hii region, respectively. 
The name and the spectral type of the stars that create these four \hii regions are taken from \cite{racine68} and \cite{downes75}.
Three are associated with reflection nebulae and are called IRS1, BD-06~1418 (vdB69), BD-06~1415 (vdB67), and HD~42004 (vdB68) 
(see Figs.~\ref{fig:hotRcoldTmaps}a, b, \ref{fig:id_temp-extraArea4prof} and Table~\ref{table_HIIchar}). 
The central UC\hii region is located at the junction of three main filaments and a few fainter ones, giving the overall impression of a `connecting hub'  (Fig.~\ref{fig:hotRcoldTmaps}b). 
The three other \hii regions are developing within this filament web and within the \uchii cloud envelope,  
which complicates study of their density structure.

The dust temperature image of Figs.~\ref{fig:hotRcoldTmaps}a and \ref{fig:id_temp-extraArea4prof} shows that the mean temperature of the cloud is about 13.5~K and that it increases up to $\sim$27~K within the three compact and ultra-compact \hii regions. 
{
Dust temperature maps only trace the colder, very large grains and are not sensitive to hotter dust present within \hii regions \citep[see][Fig.~7]{anderson2012}.  
The temperature obtained from FIR \emph{Herschel} SED corresponds to dust heated in the shell or PDR and thus directly relates to the size of the ionised bubble.
}

\subsection{Radio fluxes and spectral type of exciting stars}\label{sec:HIIRsizeYdens}

We first estimated the spectral type of exciting stars from their radio fluxes.
To do so, we looked for 21cm radio continuum flux 
 in the NVSS images  \citep{NVSS}, as well as in the corresponding catalogue.\footnote{NVSS : NRAO VLA Sky Survey, http://www.cv.nrao.edu/nvss/}
The three compact \hii regions delineated by colours and the temperature contours of Fig.~\ref{fig:hotRcoldTmaps} 
all harbour one single compact 21~cm source (see Table~\ref{table_HIIchar}). 
The more developed northern \hii region contains several 21~cm sources, whose cumulated flux can account for the total integrated emission of the region.
We calculated the associated Lyman continuum photon emission, using the NVSS 21~cm flux listed in Table~\ref{table_HIIchar} and the formulae below, taken from \cite{MartinH2005}, for example,
\begin{equation}
N_\text{Lyc}=7.6 \times 10^{46}\,\text{s}^{-1} \times \left( \frac{S_\nu}{\text{Jy}} \right) \left( \frac{T_e}{10^{4}~\text{K}} \right)^{-\frac{1}{3}} \left( \frac{d}{\text{kpc}} \right)^2 \times b(\nu,T_e)^{-1},
\label{eq:Nlyc}
\end{equation}

\noindent where ${S_\nu}$ is the integrated flux density at the radio  frequency $\nu$, ${T_e}$  is the electron temperature, $d$ is the distance to the source, and the function $b(\nu,T_e)$ is defined by
\begin{equation}
b(\nu,T_e)=1+0.3195 \log \left( \frac{T_e}{10^4~\text{K}} \right)-0.213 \log \left( \frac{\nu}{\text{GHz}} \right).
\label{eq:bNu}
\end{equation}

\citet{Quireza2006} determined an electron temperature of $\sim$8600~K
for the central UC\hii region from radio recombination lines.
We adopt the same temperature for the other \hii regions since this ${T_e}$ value is close to the mean value found for \hii regions \citep[see e.g.][Table 1]{Quireza2006}.
Using the $N_\text{Lyc}$ estimates of Table~\ref{table_HIIchar} and the calibration proposed by \citet{panagia73}, 
we assumed that a single source excites the \hii regions of Mon~R2 and estimated their spectral type (see Table~\ref{table_HIIchar}).
These radio spectral types from B2.5 to B1 agree perfectly with those deduced from the visual spectra of \citet{racine68}
\footnote
{The spectral types from \citet{HR76} were not used because they misidentified vdB~68 and vdB~69 \citep[see][note.1]{loren1977} 
and because \cite{racine68} used better spectral resolution 
}
(see Table~\ref{table_HIIchar}). 
We calculated  $N_\text{Lyc}$ for the Mon~R2 UC\hii region 
using six other radio fluxes with wavelengths from 1 to 75~cm 
available in Vizier. 
They all have radio spectral types ranging  from O9.5 to B0, which also agrees with previous estimates \citep{massi85, downes75}.
The concordance of spectral types derived from flux at different radio wavelengths, including the shortest ones,  
confirm the free-free origin of the centimetre fluxes with a limited synchrotron contamination.

\subsection{\hii region type classification}\label{sec:HIItype}

\citet[] [end of Sects.~4.a and 4.c.xi] {Wood1989} previously classified Mon R2 as an UC\hii region.
Considering the sizes and densities determined in Sect.~\ref{sec:HIIRsize},  we define the morphological type of the  four \hii regions.
The values from the central \hii region ($\rho_e = 3.2\times10^3$cm$^{-3}$,  $R_{\rm \ion{H}{ii}} = 0.09$ pc) are compatible with the range of oberved values for UC\hii regions 
\citep[e.g.][]{Hindson2012}, $\rho_e = 0.34-1.03\times10^4$ cm$^{-3}$ and $R_{\rm \ion{H}{ii}} = 0.04-0.11$ pc. It agrees with  
the classification obtained from IRAS fluxes\footnote
{The MonR2 IRAS fluxes donwloaded from Simbad ($F_{12} =470$ Jy, $F_{25} =4100$ Jy, $F_{60} =13070$ Jy, and  $F_{100} =20200$ Jy) fulfil 
the \citet{Wood1989apj} and \citet{Kurtz1994}  criteria: $log(F_{25}/F_{12})$ = 0.94 > 0.57,  $log(F_{60}/F_{12})$ = 1.44 > 1.3, and $F_{100}$ > 1000 Jy.}.
The western and eastern \hii regions have similar small sizes (around 0.1 pc) but lower densities (< 1000 cm$^{-3}$), which we assume to be more characteristic of compact \hii regions. 
The northern \hii region is larger (close to 1 pc) and has the very low density (< 100 cm$^{-3}$) typical of extended \hii regions. 

\begin{table*}  
\centering
 \caption{Properties of the four \hii region bubbles of Mon~R2}
\begin{tabular}{|c|c|ccc|ccc|cc|}
\hline   
Region name  & Type &  \multicolumn{3}{|c|}{Exciting star from visual spectra} & \multicolumn{3}{|c|}{Ionisation from 21 cm free-free}    & $R_{\rm \ion{H}{ii}}$ & $\rho_e$ \\ 
          &  &  Ident.        & Sp.T.               & $A_{V}$[mag]            &  $S_\text{21cm}$ [mJy]  &  $N_\text{Lyc}$ [s$^{-1}$]  &   Sp.T.               &  [pc]  &    [cm$^{-3}$]  \\
(1) & (2)  & (3) & (4) & (5)&  (6)  &(7)  &(8) &(9) &(10) 
\\
\hline  
central             & UC & IRS1             & -  & -  & 4114. &  $2.4\times10^{47}$          & B0  & $0.09\pm0.01$   &  3200   
 \\ 
western (vdB67)   & Compact & BD-06 1415 & B1  & 2.7 & 27.4 &  $1.6\times10^{45}$  & B1  &  $0.11\pm0.01$  &     $\sim$200
\\ 
eastern (vdB69)   & Compact & BD-06 1418 & B2.5 & 1.7 &  2.2 &  $1.3\times10^{44}$& B2.5 &  $0.08\pm0.01$  &    $\sim$100  
\\ 
northern (vdB68)   & Extended & HD 42004  & B1.5  & 1.6 &  38. &  $2.2\times10^{45}$  & B1   &  $0.8\pm0.05$  &    $\sim$10  
\\ 
\hline  
 \end{tabular}
 \label{table_HIIchar}    
 \begin{list}{}{}
\item[]
{{\bf Notes: } 
The Mon~R2 UC\hii region is also called G213.71-12.6 or  IRAS~06052-0622. 
Spectral types in Cols.~4 and 8 are given by \cite{racine68} from visual spectra and from $N_\text{Lyc}$ estimates derived from NVSS 21~cm data and Eq.~\ref{eq:Nlyc} (see Cols.~6-7).
Extinctions given in Col.~5 are estimated from E(B-V) given by \cite{HR76}.
The radii and electron density of the \hii region bubbles (Cols.~9-10) 
are estimated from the \emph{Herschel} dust temperature and column density images (see Sects.~\ref{sec:HIIRsize}-\ref{sec:HIIRdensstruct}). 
}
 \end{list}      
\vspace{-.5cm}
\end{table*}

\section{Modelling of the \hii regions' environment}
\label{s:hii_and_shell}

The initial dense cloud structures inside which the high-mass protostars have been formed  and 
the \hii regions that have developed could have been considerably modified by both the protostellar collapse and the \hii region ionisation. 
The bubble and  shell created by the expansion of \hii region retains information about its age and the mean density of the inner protostellar envelope 
whose gas has been ionised {or collected}. 
We hereafter investigate {all of the components needed to describe} the environment of \hii regions. 
The density model is presented in Sect.~\ref{s:hiiRmodel}.
We first focus on the inner components that are directly associated with \hii regions 
and which are generally studied at cm-wave radio, optical, or near-infrared wavelengths (see Sects.~\ref{sec:HIIRsize} and \ref{sec:HIIRshell}). 
We then characterise the component best studied by our \emph{Herschel} data in detail, namely the surrounding cold neutral envelope (Sect.~\ref{s:env}).
We finally jointly use all available pieces of information to perform a complete density modelling in Sect.~\ref{sec:HIIRdensstruct}.

\begin{figure}[htbp]
\vspace{-.3cm}
\includegraphics[height=.95\hsize]{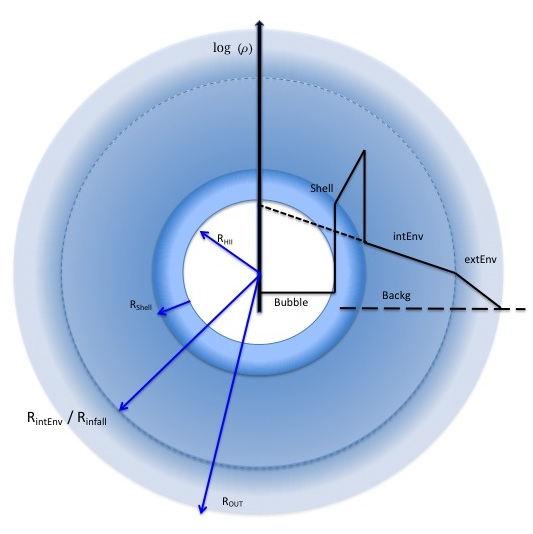}
\hfill 
\vspace{-.6cm}
\caption{ 
\hii region and its surrounding four-component spherical model and radial profile of density (right). 
The ionised bubble is first surrounded by the shell containing all of the swept-up material. 
The neutral envelope is made of two parts with different density gradients finally merging into the background. 
{The short dashed line represents the extrapolation of the inner envelope before the ionisation expansion and 
collecting process.} 
\label{fig:rhoProfilOnGeom}}
\vspace{-.6cm}
\end{figure}

\subsection{Model of \hii regions and their surroundings}\label{s:hiiRmodel}

Modelling of the observed column density and temperature structure of \hii regions and their surroundings (Fig.~\ref{fig:hotRcoldTmaps}) requires four density components (see Fig.~\ref{fig:rhoProfilOnGeom}): a 
central ionised bubble, a shell concentrating almost all of the collected molecular gas, a protostellar envelope decreasing in density, and a background.
The background corresponds to the molecular cloud inside which the \hii region is embedded.
The three other components are proper characteristics of the \hii region and its envelope.  
The envelopes of the three compact and UC \hii regions are not correctly described by a single power~law, but require two broken power~laws 
similar to that of the  inside-out collapse protostellar model \citep{Shu1977,Dalba2012,Gong2013}.

\subsection{Extent of \hii regions }\label{sec:HIIRsize}

\hii region sizes are important for measuring because they constrain
the age and evolution of the regions.  Sizes are generally
computed from H$\alpha$ or radio continuum observations that directly trace the
ionised gas.  Such data are not always available. Potentially, H$\alpha$ could be attenuated for the youngest HII regions, and radio continuum
surveys frequently have relatively low resolution.  We can instead use
\emph{Herschel} temperature maps, as well as 70\,\microm\ and 160\,\microm\ images to estimate the size of the \hii regions.

\begin{figure}[htbp]
\includegraphics[height=.75\hsize]{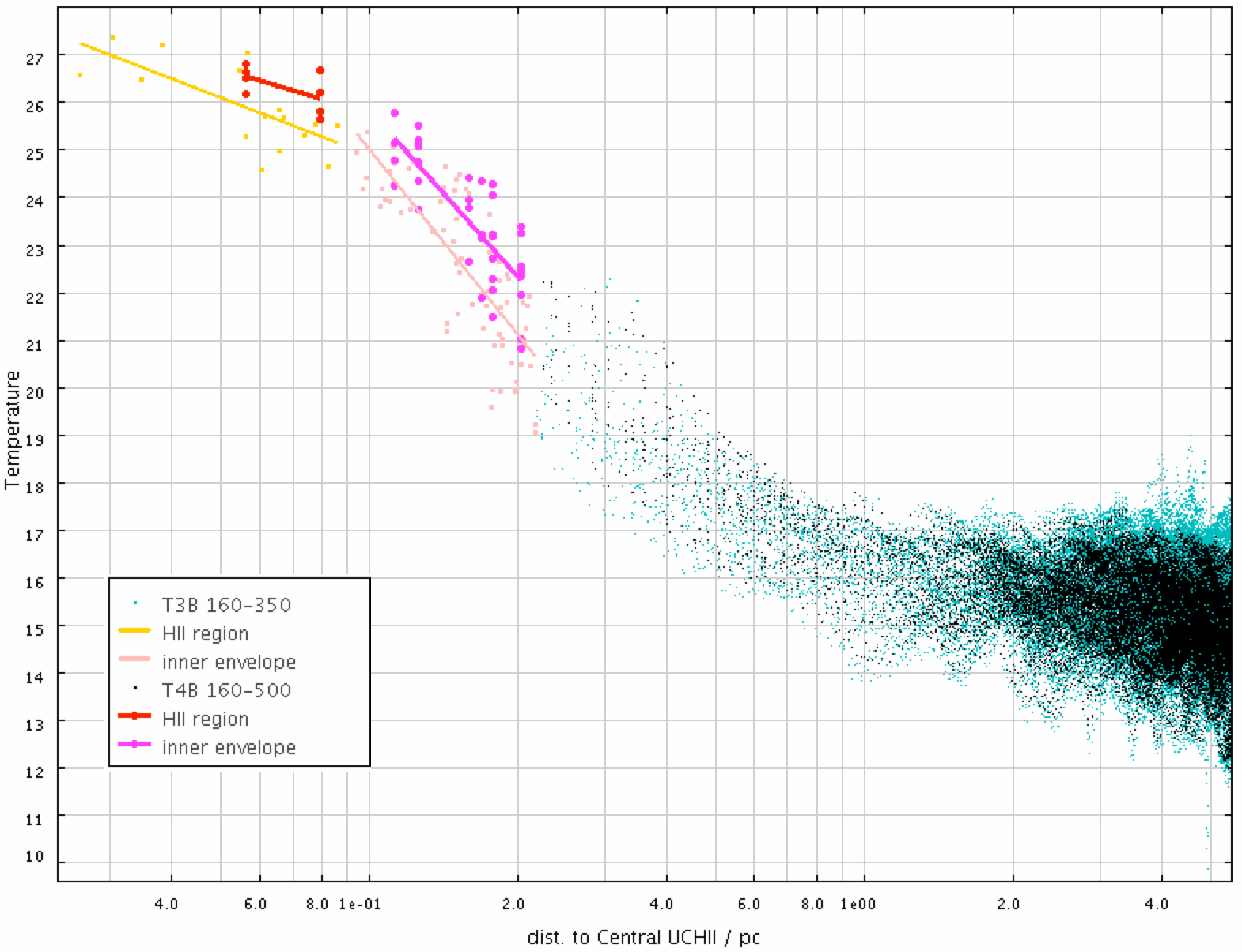}
\hfill 
\vspace{-.2cm}
\caption{Dust temperature radial profiles of the central \uchii region showing a steep outer part and a flatter inner region.
We note the agreement of the measurements done with the  3-band temperature map, $T_{\rm 3B}$ (red and magenta lines and points),   
and with the 4-band temperature map, $T_{\rm 4B}$ (yellow and pink lines and points).
\label{fig:norm4Profil4center}}
\vspace{-.3cm}
\end{figure}

The extent of all four \hii regions can clearly be seen in the dust temperature map shown in Fig.~\ref{fig:hotRcoldTmaps}a.  
This is an indirect measure since the \emph{Herschel} map traces the temperature of big dust grains that dominate in dense gas and 
which reprocess the heating of photo-dissociation regions (PDRs) associated with \hii regions.
Such a dust temperature map provides the temperature averaged over the hot bubbles in the \hii region and colder material seen along the same line of sight. 
This is particularly clear when inspecting the western \hii region, which is powered by BD-06~1415, since its temperature structure is {diluted and distorted} by the filament crossing its southern part. 
We used both the $T_{\rm 4B}$ and $T_{\rm 3B}$ temperature images with $36\arcsec$ and $25\arcsec$ resolutions. 
For the large northern  extended \hii region, the  $T_{\rm 4B}$ map alone would be sufficient since the angular resolution is not a problem. 
For the three other \hii regions, we need to use the $T_{\rm 3B}$ map, as well as 70\,\microm\ and 160\,\microm\ maps 
where intensities are used as temperature proxies (see discussion by \citealt{2011dustem,fredo2011}).
The 70 $\mu$m emission traces the hot dust and small grains in hot PDRs, which mark the borders of \hii regions and thus define their spatial extent. 

Despite the inhomogeneous and filamentary cloud environment of the Mon~R2 \hii regions, the heated bubbles have a relatively circular morphology (see Fig.~\ref{fig:hotRcoldTmaps}a).
{
This is probably due to the ionisation expansion, which easily blows away filaments and dense structures \citep[e.g.][]{minier2013}.
}
To more precisely define the \hii region sizes, we computed temperature radial profiles azimuthally averaged around the heating and ionising stars,
which are shown in Figs.~\ref{fig:hotRcoldTmaps}a-b. 
The resulting temperature profiles of the four \hii regions share the same general shape: a slowly-decreasing 
central part surrounded by an envelope where the temperature follows a more steeply decreasing power~law (see e.g. Fig.~\ref{fig:norm4Profil4center}). 
We used the intersection of these two temperature structures to measure the radius of  the flat inner part, which we use as an estimator of  the \hii region radius, $R_\text{\hii}$ (see Table~\ref{table_HIIchar}). 
At the junction between the two temperature components, data scattering and deviation from a flat inner part leads to uncertainties of about 15\%. 
The higher resolution $T_{\rm 3B}$, 70\,\microm, and 160\,\microm\ maps confirm the derived $R_{\rm \ion{H}{ii}}$ values and reach a confidence level of $\sim$10\% uncertainty. 

\begin{figure}[htbp]
\includegraphics[height=.78 \hsize]{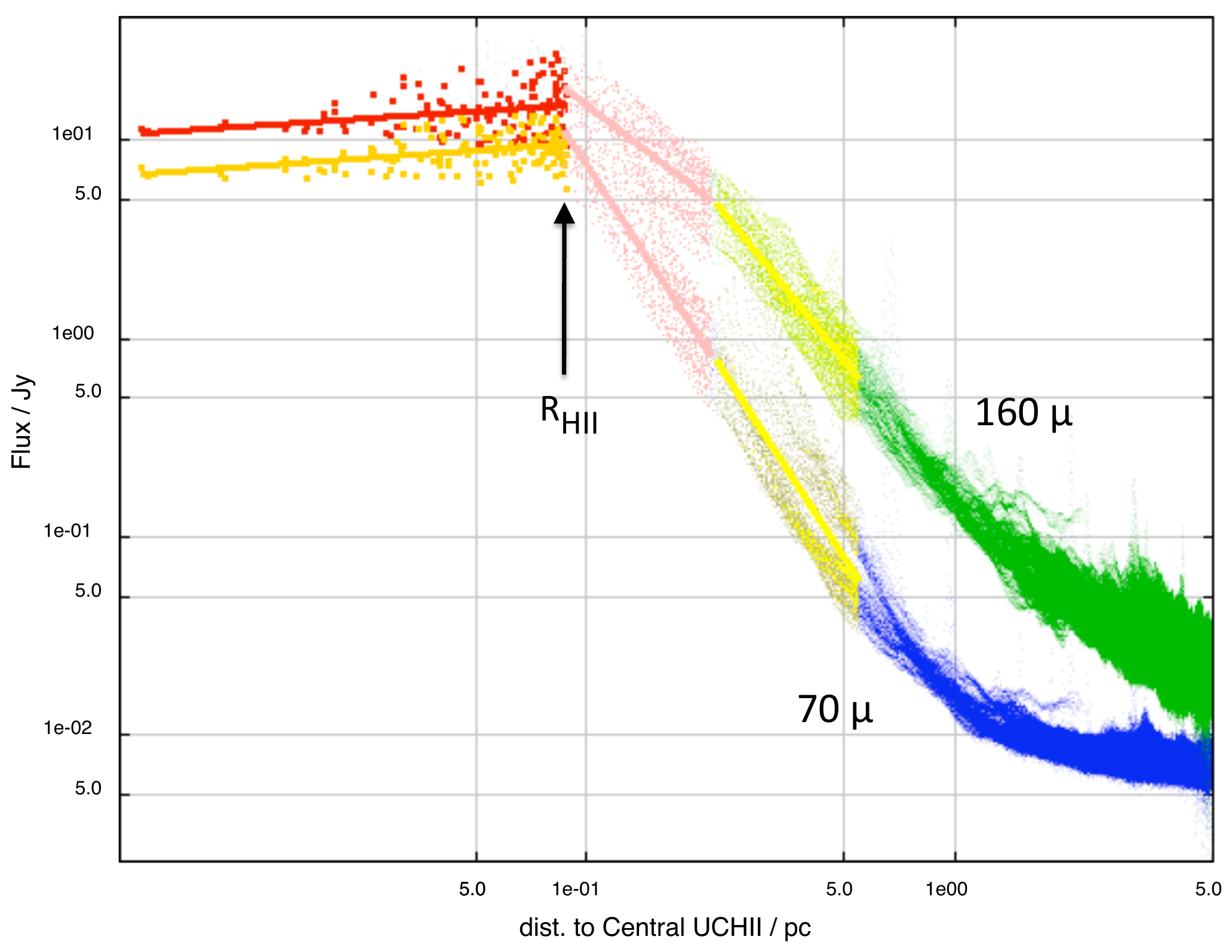}
\hfill 
\vspace{-.6cm}
\caption{Radial intensity
profiles of the central \uchii region showing the measurements done at 70\,\microm\,(blue points) and 160\,\microm\,(green points) taking advantage of the PACS highest resolution,
compared with the  3-band temperature ($T_{\rm 3B}$)  and the 4-band temperature ($T_{\rm 4B}$) data given by the black arrow. 
The 160\,\microm\ profile decreases less in the inner part of the envelope (pink line) than in its outer part (yellow line). 
Since 160\,\microm\ emission is sensitive to dust temperature and density, this behaviour is probably related to the two different slopes found for the density profile of the 
envelope  characterised in Sect.~\ref{sec:HIIRenvdensprof}. The slope change indeed occurs at similar $R_\text{Infall}$ radius. 
\label{fig:I70160Profil4center}}
\vspace{-.3cm}
\end{figure}

For the (small) central \uchii region, the radial profiles of the $T_{\rm 3B}$ and $T_{\rm 4B}$ temperatures are shown in Fig.~\ref{fig:norm4Profil4center}.
The flux profiles obtained from the 70\,\microm\ and 160\,\microm\ images 
{
with ${6}\arcsec$ and $12\arcsec$ resolutions
} 
are shown in Fig.~\ref{fig:I70160Profil4center}.
These four different estimators display similar radial shapes for the \hii region and 
give consistent size values: $R_{\rm \ion{H}{ii}} \simeq 0.09$, 0.08, 0.09, and 0.09~pc for 70\,\microm, 160\,\microm, $T_{\rm 3B}$, and $T_{\rm 4B}$, respectively. 
Similarly, concordant measurements are found for the fully resolved northern \hii region in the $T_{\rm 4B}$ and $T_{\rm 3B}$ maps: $R_{\rm \ion{H}{ii}}\simeq (0.85\pm0.05)$~pc. 
Interestingly, for the central \uchii region, the derived $R_{\rm \ion{H}{ii}}$ value of $\sim$0.09~pc is in excellent agreement 
with the one deduced from H42$\alpha$ RRL measurements \citep[$\sim$0.08~pc, see][]{pilleri2012},
showing that hot dust and ionised gas are spatially related.
{
Radio continuum data  and infrared [Ne II] line also give similar sizes and shapes \citep{massi85, Jaffe2003}. 
}

Using Lyman flux (Table~\ref{table_HIIchar} Column 7) and \hii region size (Table~\ref{table_HIIchar} Column 9) we use Eq.~\ref{exptime} to compute  the corresponding electron density $\rho_e$, given in column 10.

\subsection{\hii region 
shells and  their contribution to column density} \label{sec:HIIRshell}
 
Dense shells have been observed around evolved \hii regions within simple and low-column density background environments \citep[e.g.][]{Deharveng2005,Churchwell2007,anderson2012}.
The `collect-and-collapse' scenario proposes that the {material ionised by UV} flux of OB-type stars is indeed efficiently sweeping up 
some neutral gas {originally} located within the {present} \hii region extent and developing a shell at the periphery of \hii bubbles \citep{ElmegreenLada1977}. 

In Appendices~\ref{sec:HIIshellContribution} and \ref{sec:HIIshellAttenuation} 
we investigate the contribution of this very narrow component to the column density measured with \emph{Herschel}. 
Appendix~\ref{sec:HIIshellContribution} shows the beam dilution of the column density expected from the shell shoulder.
Appendix~\ref{sec:HIIshellAttenuation} explores the density attenuation due to bubble and shell expansion in different types of envelope density profile.
These two effects combined reduce the importance of the shell and predicts a decrease in the observed column density of \hii regions with expansion at least in decreasing-density envelopes.
The shell of the central UC\hii region characterised by \citet{Pilleri2014, Pilleri2013} has an unnoticeable contribution to the high column density of the envelope. 
Even in the northern region, which is the most extended and favourable to the detection of a shell, 
the shell characteristics are uncertain due to 
the filamentary environment inside which the \hii region developed  (see Fig.~\ref{fig:4compProfilNorth}). 
It shows that the shell is not a dominant component of the column density structure of the \hii regions studied here, but it cannot be neglected {a priori} from modelling.

\subsection{The envelope surrounding \hii regions}\label{s:env} 

\begin{figure*}[!ht]
\begin{center}
\subfloat [column density radial profile of the central UC\hii region]{\includegraphics[ scale=0.42]{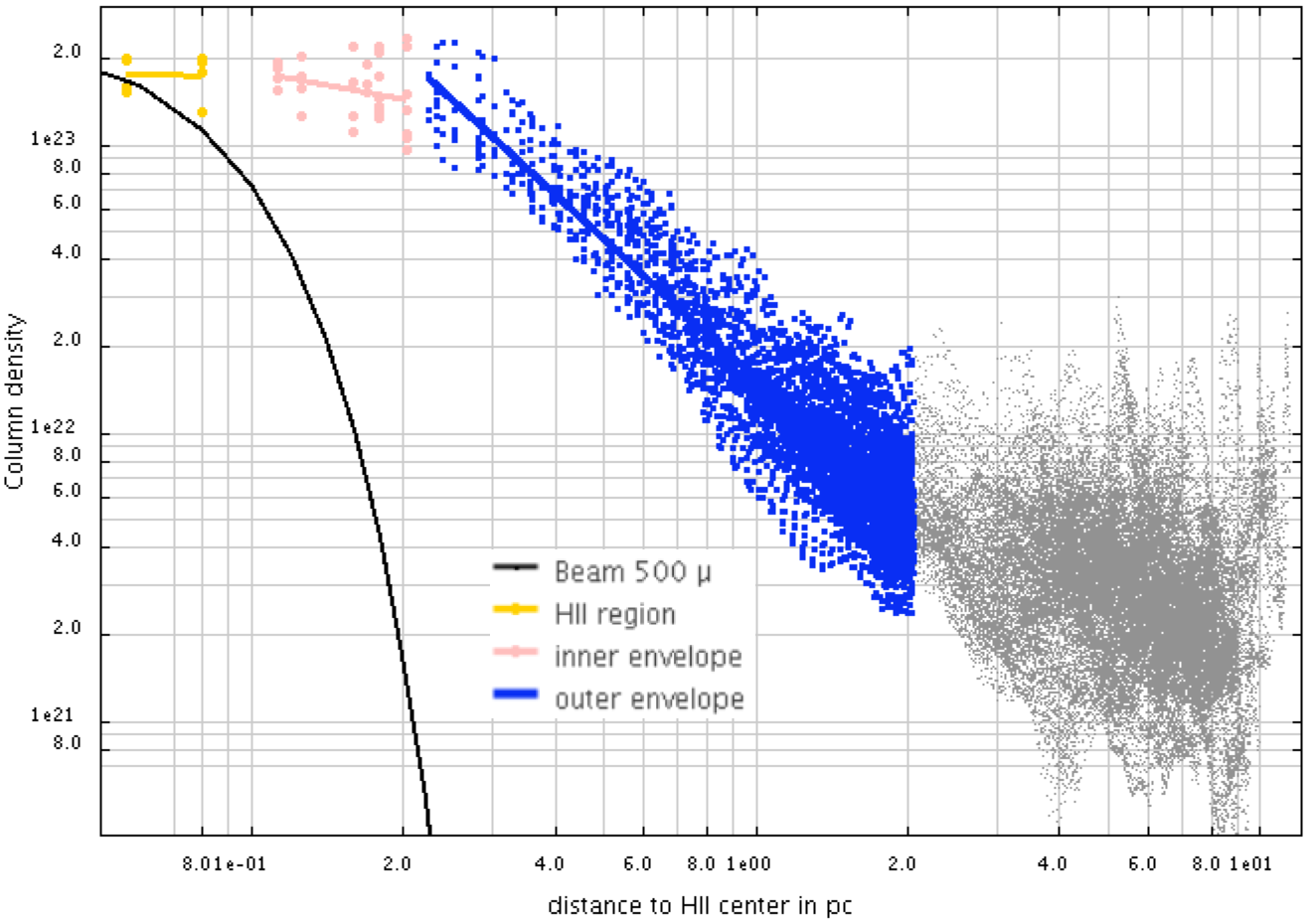}}
\subfloat [column density radial profile of the eastern C\hii region] {\includegraphics[ scale=.46]{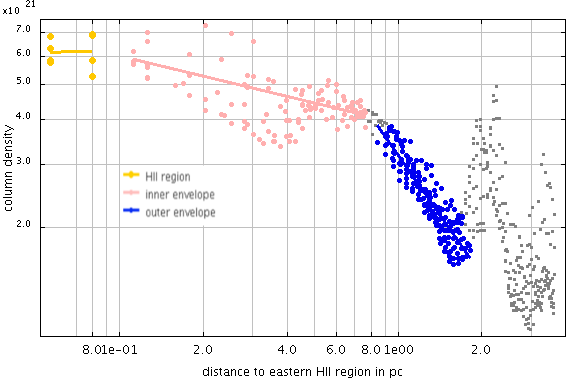}}
\vspace{-.03cm}
\caption{\hii region bubble (yellow line and points) surrounded by a neutral gas envelope splitting into slowly and sharply decreasing $N_{H_2}$ parts (pink and blue lines and points).
The background is defined by the lower limit of the clouds of grey points. The resolution corresponding to the Herschel beam at 500\microm\,is illustrated by the black curve in a)  
\label{fig:2subfigColdensProf} }
\vspace{-.5cm}
\end{center}
\end{figure*}

Amongst the crucial parameters influencing the expansion of an \hii region and the size of the heated/ionised regions, we need to characterise the density profiles of their neutral envelopes. 
We therefore constructed radial column density profiles by plotting 
the $N_{\rm H_2}$ map around each of the ionising sources shown in Fig.~\ref{fig:hotRcoldTmaps}b. 
Figures~\ref{fig:2subfigColdensProf} shows the column density profiles of the central \uchii  and eastern \hii regions, respectively. 
The colours used for the different radial parts of the profiles are similar to those used for the temperature profiles of Fig.~\ref{fig:norm4Profil4center}. 

Radial column density profiles exhibit relatively flat inner parts 
that provide a reliable measurement of the column density maximum value, $N_\text{H2}^\text{Max}$, 
of the envelope surrounding each of the four \hii regions (see Table~\ref{table_Envchar}, Col.7).
The sizes of these envelopes, given by the radius at half maximum, $R_\text{HM}$, 
were measured on the column density profile
and range from 0.35~pc to 1.5~pc    
(see Table~\ref{table_Envchar}, Col.8). 

Assuming a spherical geometry that agrees with the shape of \hii regions {studied here},  for which $R_\text{HM}$ can be used as a size estimator, 
we can roughly estimate the observed mean density, $\langle\rho_\text{obs}\rangle$, {from the maximum of the surface density $N_\text{H2}^\text{Max}$}, 
using the equation 
\begin{eqnarray}
\label{eq:meanden}
\langle\rho_\text{obs}\rangle & = & \frac{ n } { V }   
                               = \frac{ \pi\, \times R_\text{HM}^2 \,N_\text{H2}^\text{Max} } { \frac{4}{3}\,\pi \times R_\text{HM}^3 } 
                               = \frac{3}{4} \times \frac{ N_\text{H2}^\text{Max} } { R_\text{HM} }.
\end{eqnarray}
This mean density corresponds to the presently observed state of the envelope (see Table~\ref{table_Envchar}, Col.9).
To estimate the initial mean density of the envelope, $\rho_\text{initial}$, before the ionisation starts,
Sect.~\ref{sec:HIIRdensstruct} suggests another way to determine it
 through its actual maximum density and density profile slope.
 
Comparison of the central UC\hii region's characteristics shows good agreement, as deduced here with the parameters of the model used by \cite{pilleri2012}.
The model has an envelope size of 0.34~pc, inner radius of 0.08~pc, 
and a mean density $\sim0.8\times 10^5$~cm$^{-3}$ comparable to our estimates: 
$R_{\rm \ion{H}{ii}}\sim0.09$~pc $R_\text{HM} \sim 0.35$~pc,  $\langle\rho_\text{obs}\rangle = 1.4\times 10^5$~cm$^{-3}$.

\subsubsection{Column density profiles}\label{sec:HIIRenvdensprof}

With the model of Fig.~\ref{fig:rhoProfilOnGeom} in mind, we characterise the four main components associated with \hii regions and their surroundings
in detail. 
The  \hii region bubble, fitted by a yellow line, corresponds to a temperature plateau and has an almost constant column density in Fig.~\ref{fig:2subfigColdensProf}. 
Surrounding the \hii region, one can find the cloud envelope, which has a constantly decreasing temperature profile and splits here into slowly and sharply decreasing $N_{\rm H_2}$ parts,
fitted by pink and blue lines in Fig.~\ref{fig:2subfigColdensProf}. 
As for the eastern \hii region, the column density is lower and the inner envelope extent is larger, allowing a better distinction between the different envelope parts than in the central \uchii region. 
In this context, we define the inner envelope as the part where the column density slowly decreases and the external envelope as the one with a steeper decrease.
(
The crossing point between these two envelope components defines the outer radius of the inner envelope, 
called $R_\text{Infall}$ for reason explained in Sect.~\ref{sec:protinfall}. Its values are given in Table~\ref{table_Envchar}. 

The profile analysis is trustworthy for the central \uchii region since it is prominent and azimuthally averaged over $2\,\pi$ radians. 
The analysis is cruder for the three other \hii regions developing between filaments (see Fig.~\ref{fig:hotRcoldTmaps}b).  
For these \hii regions, we have selected inter-filament areas and quadrants that represent the \hii regions best and avoid the ambient filamentary structure (see Fig.~\ref{fig:id_cdens-extraArea4prof}). 
By doing so, we aimed to measure the contribution of the \hii region envelope alone. 
We have checked, mainly in the case of the eastern region, that varying the azimuthal sectors selected between the major filaments does not change either the radii 
or the slope of the envelopes by more than 20\%. 
However, contamination by other overlying cloud structures cannot be completely ruled out, and some studies of other isolated prominent \hii regions are needed to confirm the results presented here. 

The column density slopes of both the internal and the external envelopes can be represented well by power~laws such as $N(\theta)\propto \theta^{p}$
(see Fig.~\ref{fig:2subfigColdensProf}). 
Since the column density is the integration of the density on the line of sight, we can logically retrieve their density profiles through measuring the column density profiles. 
For a single power~law density $\rho(r)\propto r^{q}$, an  asymptotic  approximation leads to a very simple relation between the $q$ index and the power~law index of the column density profile described as $N(\theta)\propto \theta^{p}$: $q\simeq p-1$. 
This assumption holds for piecewise power~laws that match a large portion of the observed envelopes.  
It is the case of external envelopes as is confirmed in Sect. \ref{sec:HIIRdensstruct}.
However, for small pieces like the inner envelopes, the conversion from column density to density power~law indices is more complex 
\citep[see][]{Yun1991,Bacmann2000,Nielbock2012}. 

Towards the centre, at an impact parameter\footnote{ Impact parameter: radial distance from the line of sight to the centre.}  
smaller than the \hii region outer radius, all four density components contribute to the observed column density. 
Then a gradual increase in the impact parameter progressively  reduces the number of the contributing density components.  
The four structural components of Fig.~\ref{fig:rhoProfilOnGeom} were constrained, step by step, from the outside/background to the inside/HII region bubble. 

We first estimated the background  column density arising from the ambient cloud using both near-infrared extinction
\citep{Schneider2011} and the \emph{Herschel} column density map (Fig.\ref{fig:id_cdens-extraArea4prof}). 
They consistently give background levels of $A_{\rm v}\sim 0.5$~mag\footnote{We used the relation by \cite{bohlin78} to transform \emph{Herschel} $N_{\rm H_2}$ measurements into $A_\text{V}$ values in mag unities, $A_\text{v} \simeq 10^{21}\,\text{cm}^2 \times N_{\rm H_2}$.}
at the edge of the \emph{Herschel} map.
To minimise the systematic errors in determining the slope of the outer envelope, a 0.5~mag background was subtracted from the column density profiles. 
 For the northern  \hii region, the power~law slope measured for the envelope remains well-define with 
p possibly ranging from -0.4 to -0.5, although
the background contribution to the observed column density profile is the
most important. 

We simultaneously estimated the power~law coefficients of {the two parts of the} envelope from the slopes ($p$) measured on `{background}-corrected' column density profiles.
We used the relation $q\simeq p-1$ to measure  the density profile slopes ($q$) of the outer envelope  {and obtain a first guess for the inner envelope}.
Tables~\ref{table_Envchar} and \ref{table_EnvcharComp} list  {some} of the 
parameters derived from our {column} density structural analysis. 

\begin{table*}  
\centering
 \caption{{Comparison} of the properties of the neutral envelopes surrounding the four \hii regions of Mon R2}
\begin{tabular}{|c|cccc|c|cccc|}   
\hline  
\multicolumn{1}{|c|}{  }
& \multicolumn{4}{|c|}{ Direct fits on observed $N_\text{H2}$ profiles} 
& { Calculated }
& \multicolumn{4}{|c|}{ Fits of reconstructed $N_\text{H2}$ profiles} 
\\ 
 Region  name   & $N_\text{H2}({\rm 1pc})$      & $q_\text{out}$ & $q_\text{in}$ &  $R_\text{Infall}$ & 
$\rho_\text{1pc}$ &
$\rho_{\rm1pc}$ & $q_\text{out}$ & $q_\text{in}$ &  $R_\text{Infall}$ \\
  &  [cm$^{-2}$]        &  &   & [pc] & 
  [cm$^{-3}$] &
  [cm$^{-3}$] & & & [pc]      \\  
 (1) & (2) & (3) & (4) & (5) & (6) & (7) & (8) &(9) & (10) \\ \hline  
central &  
$1.5\times10^{22}$  & $-2.7 $ & $-1.3  $ &   $0.25\pm.05$        & $2000.\pm200.$ &
$1900\pm250$  & $-2.75 $ &  $-0.85\pm.35$ &$0.3\pm.1$  \\ 
western &
$3.5 \times10^{21}$  & $-1.8$ & $-1.55  $ &   $0.5\pm.1$   & $300.\pm50.$ &
$370\pm50$  & $-1.9 $ &  $-1.45\pm(.1)$ &$0.5\pm.1$  \\ 
eastern &
$4. \times10^{21}$  & $-2.2$ & $-1.3  $ &   $0.8\pm.2$   & $470.\pm50.$ &
$430\pm50$  & $-2.45 $ &  $-.4\pm0.2$ &$1.\pm.2$  \\ 
northern &
$2. \times10^{21}$  & \multicolumn{2} {c}{$-1.5  $ }  &  -    & $125.\pm25.$ & $105\pm30$  & \multicolumn{2} {c}{$-1.45\pm.15$} & -  \\ 
\hline  
 \end{tabular}
 \label{table_EnvcharComp}    
 \begin{list}{}{}
\item[]
{{\bf Notes:} 
Cols.~2-5 are directly fitted from the observed  column density profiles, while Cols.~7-10  are adjusted from modelled column density profiles reconstructed from fixed density profiles.
Density profile indices and infall radius Cols.3-5 and Cols.8-10 are defined in Sect.~\ref{sec:HIIRenvdensprof}-\ref{sec:HIIRdensstruct} and Fig.~\ref{fig:rhoProfilOnGeom}.
The envelope density at 1 pc of Col.~6 is calculated with Eqs.\ref{rho1calc} and \ref{zetaFormula} from Cols.2-5 data, while the value given in Col.~7 is taken from the density model.
}
 \end{list}      
\vspace{-.5cm}      
\end{table*}
 
The outer envelope of the two compact eastern and western \hii regions displays a column density index of  
$p_\text{out} \simeq -1$ with an uncertainty of 0.2 ($\sim$$3\sigma$), which corresponds to a $\rho(r)\propto r^{-2}$ density law. 
The central \uchii region itself exhibits a steeper density gradient in its outer envelope with $q_\text{out}\simeq-2.5$. 
In practice, the uncertainties are such that the slopes are well defined within {an uncertaintiy of} about 10\% or $\sim$0.2, clearly allowing the distinction between $\rho(r)\propto r^{-2}$ and $\rho(r)\propto r^{-1.5}$ or $\rho(r)\propto r^{-2.5}$ power~laws.
This is a clear improvement from the studies made before \emph{Herschel} from single band sub-millimetre observations \citep[e.g.][]{Motte2001,Beuther2002,Mueller2002}.

The northern \hii region, whose envelope does not split into an inner and outer parts, has a decreasing density profile up to 0.8~pc with a power~law coefficient of $q \simeq-1.5$.

The internal envelopes of the central, eastern, and western \hii regions have a {column} density profile with power~law coefficients $p_\text{in} \simeq -0.35\pm0.2$, 
but the conversion is not straigthforward and the density modelling made in Sect.~\ref{sec:HIIRdensstruct} is needed to derived correct values. 

\subsection{Density modelling}\label{sec:HIIRdensstruct}

\begin{figure}[htbp]
\includegraphics[height=.72\hsize]{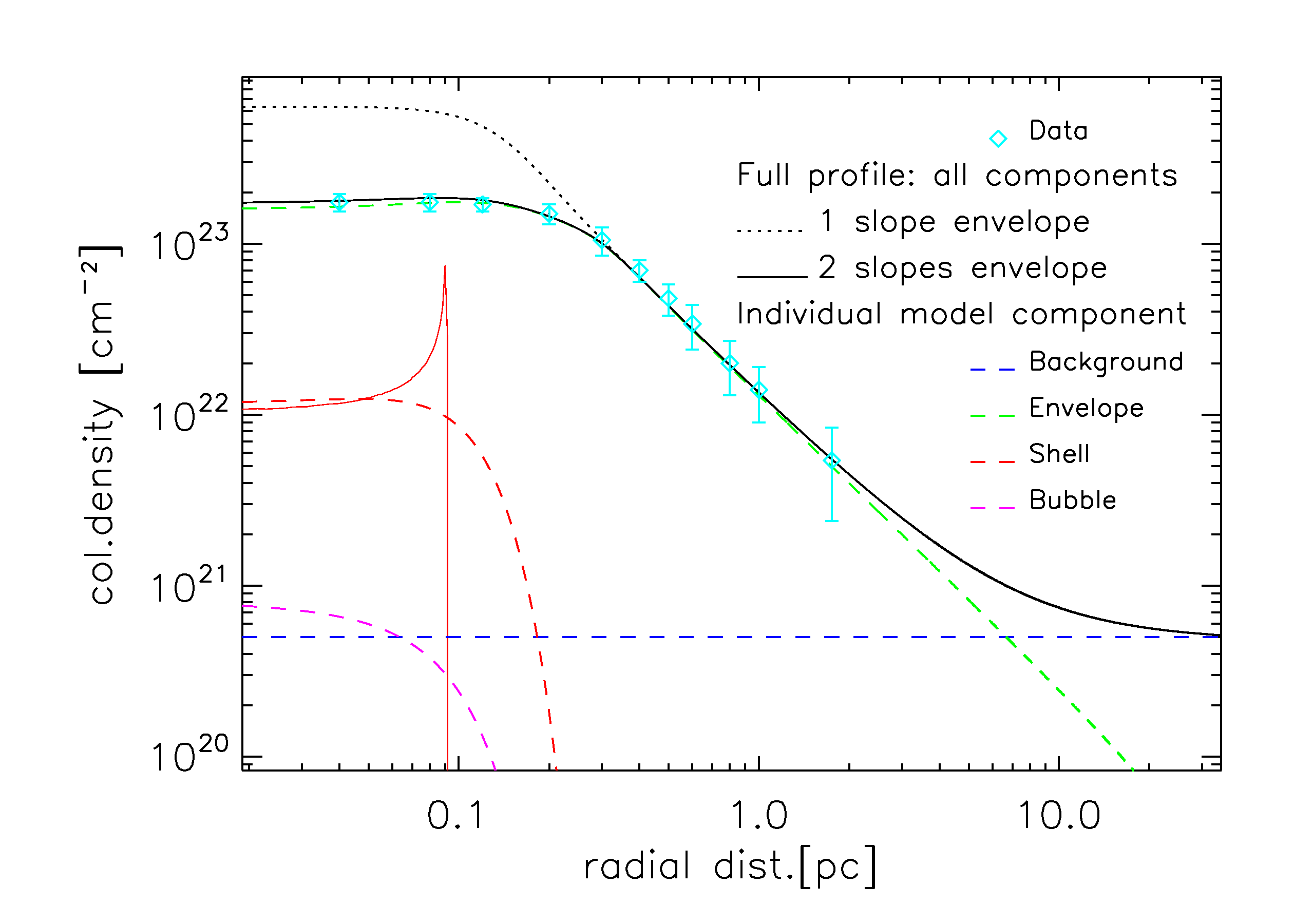}
\hfill 
\vspace{-.9cm}
\caption{Column density profile of the central \uchii region {(cyan diamonds with $\sigma$ errors bars)} compared to models with all components illustrated in Fig.~\ref{fig:rhoProfilOnGeom}. 
The neutral envelope with two density gradients or with a single one are shown by a black line and a dotted black line respectively.
An envelope model with a single density gradient cannot match the observed data.
Individual components after convolution by the beam are shown by magenta, red, green, and blue dashed lines, for the bubble, shell, envelope and background..
The continuous red line shows the unconvolved shell profile at an infinite resolution.
\label{fig:4compProfilCenter}}
\vspace{-.3cm}
\end{figure}

\begin{figure}[htbp]
\includegraphics[height=.73\hsize]{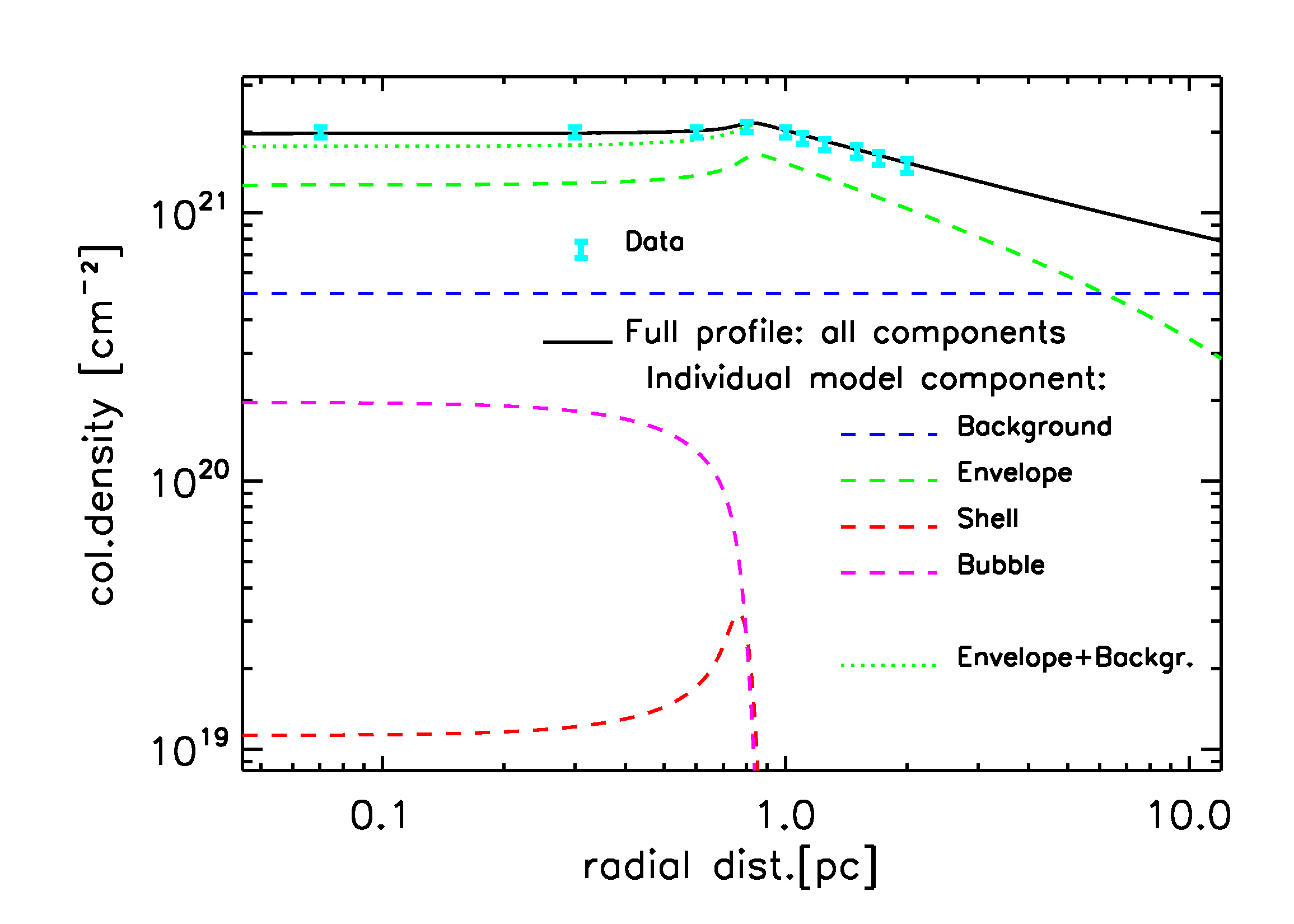}
\hfill 
\vspace{-.9cm}
\caption{Column density profile for northern \hii region and its four components coded with the same colours as in Fig.~\ref{fig:4compProfilCenter}. 
The dotted green line corresponding to the sum of the envelope and background contributions matches well with observed data.
\label{fig:4compProfilNorth}}
\vspace{-.4cm}
\end{figure}

We here perform a complete modelling of the observed column density profiles to adjust and evaluate the relative strength 
{and the characteristics} of the components used to describe \hii regions and their surroundings.
We modelled four different components along the line of sight to reconstruct the observed column density profile. 
We used the size and density profiles derived in previous sections and listed in Table~\ref{table_Envchar} to numerically integrate between proper limits, 
along different lines of sight, the radial density profiles of the background, envelope (external then internal), \hii region shell, and bubble. 
For a direct comparison with observed profiles, we convolved the modeled column density profiles with the 36\arcsec~resolution of the \herschel~$N_{\rm H2}$ maps.
Figures~\ref{fig:4compProfilCenter}-\ref{fig:4compProfilNorth} display the calculated column density profiles of each of the four aforementioned structural components 
and the resulting cumulative profile.
Figure~\ref{fig:id_cdens-extraArea4prof} also locates, with concentric circles, the different components used to constrain the structure of the central \uchii region.

\begin{table*}  
\centering
 \caption{Properties of the neutral envelopes surrounding the four \hii regions of Mon R2}
\begin{tabular}{|c|cccc|c|ccc|cc|}   
\hline   
Region name     & $R_\text{Infall}$     & $R_\text{out}$        & \multicolumn{2}{c|}{$\rho(r)\propto r^{q}$} &Infall & 
$N_\text{H2}^\text{Max}$ &  $R_\text{HM}$ & $\langle\rho_\text{obs}\rangle$ &
$ \rho_\text{env}(\rm1pc)$ & $\rho_\text{env}(R_{\rm \ion{H}{ii}})$ \\ 
  &  [pc]                       &[pc] & $q_\text{in}$ & $q_\text{out}$ & age [Myr] &   
        [cm$^{-2}$] &  [pc]      &   [cm$^{-3}$]   &     
[cm$^{-3}$]     &[cm$^{-3}$] \\ 
 (1) & (2) & (3) & (4) & (5) & (6) & (7) & (8) &(9) & (10)  &(11) \\ 
\hline  
central &  $.3\pm.1$  & $2.5\pm.5$ & $-.85\pm.35 $ & $-2.7$ & $.5-1.5 $ & 
$2.\times10^{23}$  & 0.35        & $1.4\times10^{5}$ &
$2000\pm300$  & $1.5\times 10^5$  \\  
western &  $.5\pm.1$ & $3.\pm1.$ & $-1.5\pm.15 $ & $-1.8$ & ${.8-}2.5 $ & 
$16.\times10^{21}$  & 0.4        & $9.\times10^{3} $ & 
$350\pm50$   & $12000$ \\ 
eastern  &  $.9\pm.2$ & $2.5\pm1.$ & $-.4\pm.2 $ & $-2.3$ & ${1.4-}4.4 $ & 
$7.\times10^{21}$  & 0.9          & $1.6\times10^{3} $ &
$450\pm50$   &  $1200$ \\  
northern &  {$> 2$} & $2.5\pm.5$ & $-1.45\pm.15 $ & - & $>{3.-}10. $ & 
$3.\times10^{21}$  & 1.5        & $300.$ &
$115\pm30$ & $150$ \\ 
\hline  
 \end{tabular}
 \label{table_Envchar}    
\vspace{-.2cm}
 \begin{list}{}{}
\item[]
{{\bf Notes:} 
{Cols.~2-5} describe the density profile of the neutral envelopes surrounding \hii regions by listing the radius separating the inner and outer envelope components, the outer radius, and the density power~law slopes of the two envelope components. 
 Uncertainties (3$\sigma$) on $q_\text{out}$ are  estimated to be 0.2. 
{Col.~6} gives the time elapsed since the beginning of infall and protostellar collapse. 
It is calculated from $R_\text{Infall}$ (Col.~2) and Eq.~\ref{inftime} with a velocity range from one to three times the sound speed  
(see Sect.~\ref{sec:protinfall}) {and has a statistical error of 20\%}. 
{Cols.~10-11} give the absolute values of the density measured at $r=1$~pc and at the $R_{\rm \ion{H}{ii}}$ radius, from the complete modelling of column density profiles   
(see Sect.~\ref{sec:HIIRdensstruct}). 
Col.~11 is used to estimate, through Eq.~\ref{rhoinit}, the mean initial neutral gas density before the ignition of an \hii region.
}
 \end{list}      
\vspace{-.6cm}   
\end{table*}

We recall that the background is itself defined as a constant-density plateau: $A_{\rm V}\sim 0.5$~mag for all \hii regions. 
The background, even if faint, has an important relative contribution towards the outer part of the \hii region structure, 
where the envelope reaches low values. 
For the northern  \hii region, the background contribution to the observed column density profile is  almost equal to that of the envelope, 
as it can be seen on the reconstructed profile of Fig.~\ref{fig:4compProfilNorth}. 

Our goal is to estimate for each of the four \hii regions in Mon~R2, the absolute densities of the envelope, shell, and \hii region bubble 
(see Tables~\ref{table_HIIchar}-\ref{table_Envchar}).

We adjusted our model to the complete column density profile, progressively adding the necessary components : outer envelope, inner envelope, shell and bubble,   
the latter first assumed to be empty. In Table  \ref{table_EnvcharComp}  we compare the results directly obtained from observationnal data and from data modelling.

We first modeled the density profile observed at radii greater than 1~pc, dominated by the outer envelope density. 
We jointly adjusted the density at 1 pc, $\rho_1$ and the density power~law slope, $q_{out}$.
The $q_{out}$ values derived from the $q = p-1$ relation are in excellent agreement with the fitted ones. 
The $\rho_1$ values obtained are very similar to those calculated
with Eqs. \ref{rho1calc} and \ref{zetaFormula} from the column density values at 1 pc and the appropriate $q_{out}$ values.
Our results prove that  the asymptotic assumption is correct for the outer envelope component.
The characteristics of the inner envelope, which are mixed on the line of sight with those of the outer envelope, is only reachable through modelling.
Figures~\ref{fig:4compProfilCenter}-\ref{fig:4compProfilNorth} show that once the envelope density 
{structure} is fixed, the modelled column density profile fits the observed one for radii larger than $R_{\rm \ion{H}{ii}}$ nicely. 

The shell and the bubble are expected to have a weak influence on the profiles of the ultra-compact-to-compact \hii regions (here the central, eastern, and western regions). 
Indeed, the \hii bubble sizes and the amount of gas mass collected in the shell are so small 
that their contribution can be almost neglected, and high resolution would be needed to observe the narrow 
shoulder enhancement at the border of the shell (see Fig.~\ref{fig:shellProfilCenter}). 
These two components are, however, definitely needed to reproduce the central column density of the more extended and diffuse, northern Extended \hii region. 
Figure~\ref{fig:4compProfilNorth} displays an inner column density plateau and a tentative shoulder of the column density profile just at the $R_{\rm \ion{H}{ii}}$ location. 
We have estimated an upper limit to the mean density, column density, and mass for the northern \hii region swept-up shell, assuming a 0.01~pc thickness with a density varying from 50 to 1000~cm$^{-3}$:  $N_\text{H2}^\text{shell}\sim1.6\times 10^{19}$~cm$^{-2}$, and $3~M_\odot$.
This putative shell is, however, difficult to disentangle from the envelope border and the complex cloud structure in this area. 

According to \cite{HosoInu2005}, the central bubble should be devoid of cold dust in the earliest phase of \hii region development, {roughly corresponding to the compact \hii phase.}
In the case of the central \uchii region, we used the electron density of $1.6\times 10^3$~cm$^{-3}$ \citep{Quireza2006} as an upper limit for the \hii bubble density of neutral gas. 
Even this relatively high density value results in a negligible contribution to the modelled column density profile (see Fig.~\ref{fig:4compProfilCenter}). 
For the northern \hii region, a low-density ($\sim$50~cm$^{-3}$) 
central component needs to be introduced in the bubble to reconcile the modeled profile and the observational data. 
Without it, we would observe a relative hole or depletion in the centre of the column density profile (see red and green lines in Fig.~\ref{fig:4compProfilNorth} compared to cyan or black ones).
{The uncertainties are such that the exact value of the inner density component of the northern \hii region is not known better than to a factor of ten.  
Therefore, our modelling can only give upper-limit densities for the bubble gas content and \hii region shell.}

Table~\ref{table_Envchar}  lists the adopted values of the envelopes density characteristics. 
The envelope density at 1 pc radius, i.e. $\rho_\text{env}$(r=1\,pc), which was used to scale the parametrised density profile, {defines the absolute level of density in the envelope}. 
It also gives the density observed at the inner border of the envelope, i.e. $\rho_\text{env}(r=R_{\rm \ion{H}{ii}})$, corresponding to the maximum value measured now in the envelope. 
{The central density of the eastern Compact \hii region is a bit weak and atypical.}
For the northern \hii region, the inner border of the envelope is at $R_{\rm \ion{H}{ii}}\sim 0.8$~pc, and we can
extrapolate the $\rho(r)\propto r^{-1.5}$ density profile to a distance of 0.1~pc: $\rho_\text{estimated}$(r=0.1\,pc)$\,\sim 6.6\times 10^3$~cm$^{-3}$.
We also considered the density at much smaller radii, which would correspond to the inner radius of the {extrapolated} protostellar envelope {and the central core density}:
$\rho_\text{env}(r=0.005\rm\,pc$ / 1000~AU), Col.~4 in Table~\ref{table_ionitimes}. 

All inner envelopes have a power law slope that is smaller than or equal to the expected infall value of 1.5. The values lower than 1.5 could be due to sub-fragmentation 
in the inner envelope. Some indications of inhomogeneous clumpy structures are indeed revealed by high-resolution maps (Rayner et al, in prep.).
All outer envelopes have power law slopes greater than or equal to the expected SIS value of 2, corresponding to hydrodynamical equilibrium. 
The high value observed for the central \hii region could be due to additional compresion (Sect.~\ref{sec:globalinfall}, App.~\ref{sec:griedntandcompression}). 
All these density {determination} will be used below
to estimate the mean initial neutral gas density before the ignition of an \hii region.

\section{Ionisation expansion time}\label{sec:ionisationExpansion}
 
 The physical characteristics presented in the previous sections are used here to constrain the expansion time of the ionisation bubble in its neutral envelope.
 In Sect.~\ref{sec:HIIRageCalc}, analytical calculations without gravity show that the average initial density, $\langle\rho_\text{initial}\rangle$, 
 is the main parameter dictating the expansion behaviour.
 In Sect.~\ref{sec:simus}, simulations are compared to analytical results of Sect.~\ref{sec:HIIRageCalc} to determine the impact of gravity on expansion. 
In constant high-density envelopes, gravity prevents expansion through quenching or re-collapse, but 
{in a decreasing-density envelope,} the impact of gravity becomes negligible once the expansion has started.
We thus conclude in Sect.~\ref{sec:gravDensApprox} that analytical calculations without gravity can be used to derive valid ionisation expansion times. 

\subsection{Str\"omgren-sphere expansion analytical calculations}\label{sec:HIIRageCalc}
 
Differences in size between the northern and the western \hii regions, which harbour an exciting star of the same spectral type, 
tend to suggest that the northern \hii region is more evolved than the western one (see Fig.~\ref{fig:hotRcoldTmaps}a and Table~\ref{table_HIIchar}). 
The larger number of striations observed in the {visible image of the} northern reflection nebula also indicate an older stage of evolution \citep{Thronson1980,loren1977}.

To confirm these qualitative statements, we here determine {time differences in} the beginning of the ionisation expansion \hii regions {for} the four \hii regions.
We recall that the deeply embedded central \uchii region has no optical counterpart, 
while the western, eastern, and northern  \hii regions are respectively associated with the vdB67, vdB69, and vdB68 nebulae.
\subsubsection{Average initial density}\label{sec:HIIRaverageInitialDensity}

The density observed at the inner border of the envelope (i.e. $\rho_\text{env}(r=R_{\rm \ion{H}{ii}})$, corresponding to the maximum value measured now in the envelope)  
can be used to estimate the density the protostellar envelope had at the time of the ignition of the ionisation, 
when averaged within a sphere of $R_{\rm \ion{H}{ii}}$ radius. 
Given the inner density structure observed for the four \hii regions envelopes, 
{whose density gradient is described by $\rho(r)\propto r^{q _\text{in}}$},
the mean density of the initial envelope part, which is now mostly collected in the shell, is given by 
\begin{equation}  
\label{rhoinit}
\langle\rho_\text{initial}\rangle_{R_{\rm \ion{H}{ii}}}= \left( \frac{3}{3-q_\text{in}} \right)  \times \rho_\text{env}(r=R_{\rm \ion{H} {ii}})\end{equation}
and derived in Appendix~\ref{sec:massAverDens}.
It is the most important parameter for analytical calculations,
and its calculated values are given in Col.~2 of Table~\ref{table_ionitimes}. 

\subsubsection{Expansion in an homogenous density {static} envelope}\label{sec:HIIRageCalcCstEnv}
 
\citet{spitzer78book} predicted the expansion law of a Str\"omgren sphere and thus the evolution of the \hii region size with time in an homegeneous medium (see also \citealt{dysonbook} and  \citealt[Eq.~1]{arthur2011}). 
Following \citet{ji2012}, we inverted the relation and expressed time since the ignition of the ionisation as a function of  the observed \hii region radius, $R_{\rm \ion{H}{ii}}$:
\begin{equation}
\label{exptime}
t_\text{Spitzer}(R_{\rm \ion{H}{ii}}, \langle\rho_\text{initial}\rangle)=  \frac{4}{7} \times \frac{R_{\rm Str}}{c_s} \left[  \left( \frac{R_{\rm \ion{H}{ii}}}{R_{\rm Str}} \right)^{7/4}  - 1 \right] ,
\end{equation}
where $c_s = 10$~km\,s$^{-1}$ is the typical sound speed in ionised gas, and $R_{\rm Str}$ is the radius of the initial Str\"omgren sphere \citep{Stromgren1939,Stromgren1948}. 
The latter is given by the following equation 
\begin{equation} 
R_{\rm Str} =  \left( \frac {3\, N_\text{Lyc}} {4\pi\,\alpha_\text{B} \langle\rho_\text{initial}\rangle^2 }   \right)^{1/3} ,
\label{eq_rstromgren0}
\end{equation}
or in numerical terms
\begin{equation} 
R_{\rm Str} \simeq 301\,{\rm AU}~ \left( \frac {N_\text{Lyc}} {10^{47}\rm s^{-1}} \right)^{1/3} \left( \frac{\langle\rho_\text{initial}\rangle} {10^6\rm\,cm^{-3}}  \right)^{-2/3} 
\label{eq_rstromgren0num}
\end{equation}
where $\langle\rho_\text{initial}\rangle$ is the initial mean density in which the ionisation occurs,
$N_\text{Lyc}$ is the number of hydrogen ionising photons from Lyman continuum 
and $\alpha_\text{B} = 2.6 \times 10^{-13}$~cm$^3\,$s$^{-1}$ is the hydrogen recombination coefficient to all levels above the ground.

The ionisation ages determined from Eq.~\ref{exptime} strongly depend on the 
mean initial density, $\langle\rho_\text{initial}\rangle$  (see its definition in Sect.~\ref{sec:HIIRaverageInitialDensity}),
since it defines the initial radius of the ionised sphere, $R_{\rm Str}$ (see Eq.~\ref{eq_rstromgren0}).
With the assumption that the expansion develops in a constant-density medium, the ionisation age is strictly determined by the initial and 
present sizes of the ionised region, $R_{\rm Str}$ and $R_{\rm \ion{H}{ii}}$ (see Eq.~\ref{exptime}).

Initial density estimates are very uncertain since they depend on assumptions made for the gas distribution before protostellar collapse and its evolution conditions.
Nevertheless, the envelope density structure can be approximated by the mean density of the envelope part, 
which was travelled through by the ionisation front expansion and has been collected in the shell.

Considering the maximum density of the different \hii region envelopes at their inner border ($R_{\rm \ion{H}{ii}}$) and using Eq.~\ref{rhoinit},
we calculate an initial mean density ($\langle\rho_\text{initial}\rangle_{\rm \ion{H}{ii}}$)  
reported in Col.~2 of Table~\ref{table_ionitimes}. 
Assumed to be representative of an envelope with an equivalent homogenous constant density, this value is jointly used with the ionising flux given in Table~1 to calculate the corresponding expansion time with Eq.~\ref{exptime} and is given in Col.~3 of Table~\ref{table_ionitimes}. 
The expansion time of compact and UC  \hii regions are similar ($t_\text{Exp}\sim 20-90\times10^3$~yr) with a range that is mainly due to different initial densities.
The northern extended \hii region is three to 15 times older ($\sim$300$\times 10^3$~yr), in line with what is expected from their size differences.

\begin{table*}  
\centering
 \caption{Estimations of expansion time for the four \hii regions of Mon R2}
\begin{tabular}{|c|cc|c|cc|cc|c|c|}   
\hline   
& \multicolumn{5}{|c|}{ Analytical calculations and simulations without gravity }
& \multicolumn{3}{|c|}{ Simulations with gravity }
& { Adopted }
\\
& \multicolumn{2}{|c|}{ Constant density }
& \multicolumn{3}{|c|}{ Decreasing density}
& \multicolumn{2}{|c|}{ Constant density }
&{ Dec.density} 
& { expansion } 
\\
Region & $\langle\rho_\text{initial}\rangle$ & $t_\text{Spitzer}$ &
$\rho_\text{c}$ & 
$t_\text{exp}$ (calc)&
$t_\text{exp}$ (simu.) & 
$\langle\rho_\text{initial}\rangle_\text{Max}$ & $t_\text{exp}$ (simu) &
$t_\text{exp}$ (simu) &
time \\ 
  &   [cm$^{-3}$]   &   [kyr] &
   [cm$^{-3}$]   & [kyr] &   [kyr] & 
        [cm$^{-3}$]     & [kyr] & [kyr]  & [kyr]        \\  
 (1) & (2) & (3) & (4) & (5) & (6) & (7) & (8)  & (9) & (10)\\
\hline  
central  & $2\times10^5$ & 54 & $2\times10^6$ &58 & 53 &
  $1.7\times10^5$   &  133 &  -   &   {90 $\pm$ 40} \\ 
western 
 & $2.4\times10^4$ & 92 & 
 $1\times10^6$ &108 & 98  &
 $1.7\times10^4$  
 & 215 & -
  &  {$150 \pm 50$}  \\ 
eastern  
&$1.5\times10^3$ & 23 & 
 $4\times10^3$ &24 & 23  &  
 $8\times10^3$ 
 & 170 
 & 24
  &   {25 $\pm$ 5} \\ 
northern 
& $3\times10^2$ & 310 & 
$2.5\times10^5$ &370 & 355  &
$5\times10^3$ 
 & 3700.
 & 370
  &   {350 $\pm$ 50} \\ 
\hline  
 \end{tabular}
 \label{table_ionitimes}    
 \begin{list}{}{}
\item[]
{ (2) average initial density derived from Eqs.~\ref{rhoinit} and \ref{2rho1}.\\
(3) expansion time calculated from Eqs.~\ref{exptime} and  \ref{eq_rstromgren0} and using the average density of Col.~2. \\
(4) constant density of central core derived from Eq.~\ref{rhoprof} with $\rho_\text{c} = 0.005$ pc.\\
(5) expansion time calculated from Eqs.~\ref{exptimeDecrEnv} and  \ref{eq_rstromgren0} and using the central density of Col.~4. \\
(6) expansion time derived from simulations without gravity using a decreasing-density envelope extrapolated up to the central density (Col.~4). 
Considered as a lower limit (see introduction  of Sect. \ref{sec:simus}).\\
(7) maximum constant density allowing ionisation expansion in simulations with gravity, (see Sect. \ref{sec:HIIsimuCstEnvWithG}).\\
(8) expansion time calculated by simulations using the maximum constant density of Col.~7. Considered as an upper limit.\\
(9) expansion time derived from simulations with gravity using a decreasing-density envelope extrapolated  up to highest possible central density (see Sect.~\ref{sec:HIIsimuNorthEnvWithG}).  \\
(10) adopted expansion time (see Sect. \ref{sec:HIIsimuConclusion}  and \ref{sec:gravDensApprox}).
 Errors do not take any systematic effects into account, such as those due to small-scale non-3D geometry or density inhomogeneity. \\
 }
\end{list}      
\vspace{-.7cm}    
\end{table*}

\subsubsection{Expansion in a density decreasing {static} envelope}\label{sec:HIIRageCalcDecrEnv}

The present study shows that \hii regions develop into protostellar envelopes whose density decreases with radius \citep[see Sect.~\ref{sec:HIIRdensstruct}, see also introduction in][]{Immer2014}.
The equation given by \citet{spitzer78book} describing their expansion in a medium of homogeneous density (Eq.~\ref{exptime}) therefore does not directly apply. 
We calculated the expansion of an ionised bubble in an envelope with a density profile  following a power~law.
We included in this envelope a central core with a $r_\text{c}$ radius and a constant density of $\rho_\text{c}$. 
The clump density structure thus follows
\begin{equation}  
\label{rhoprof}
\rho =  \left\{  \begin{array}{rrr} 
      \rho_\text{c} & \mbox{for} & r \leq r_\text{c} \\
      \rho_\text{1} \times r^{-q} & \mbox{for} & r > r_\text{c} \\
      &\mbox{with} & \rho_\text{c} = \rho_\text{1} \times r_\text{c}^{-q} 
   \end{array} \right .
\end{equation}
where $\rho_1$ is the density at 1~pc deduced from the constructed density profile or directly measured from the column density at the same radius (see Sect.~\ref{sec:HIIRdensstruct}).
The expansion as a function of time in a decreasing envelope is derived by calculations detailed in Appendix.\ref{sec:ionExpCalculation}. 
In close agreement with previous work by \citet[][]{Franco1990}, it is given by 
\begin{eqnarray}\
\label{exptimeDecrEnv}
t_q(R_{\rm \ion{H}{ii}}) & =  &\frac{4}{7 - 2q} \times \frac{r_\text{c}^{7/4}} {{R_{\rm Str}}^{3/4}}   \frac{1}{c_s} \left[  \left( \frac{R_{\rm \ion{H}{ii}}}{r_{\rm c}} \right)^{{7-2q}/4}  - 1 \right]\\
                                   & = & \frac{4}{7 - 2q} \times \frac{r_\text{c}} {V(r_c)}  \left[  \left( \frac{R_{\rm \ion{H}{ii}}}{r_{\rm c}} \right)^{{7-2q}/4}  - 1 \right] \nonumber
\end{eqnarray}
with the same definitions as in
Eqs.~\ref{exptime}-\ref{eq_rstromgren0} and where $V(r_c) = { \left({R_{\rm Str}/{r_c}}\right)^{3/4} } \times  {c_s}$ 
is the shell expansion speed at the homogeneous core radius $r_c$.
For a constant density, which is the Spitzer case, $q=0$ and
${r_c} = {R_{\rm Str}}$, and Eq.~\ref{exptimeDecrEnv} resumes to Eq.~\ref{exptime}.
 
The inner envelope density profiles  of the four regions {have gradients which  at most} corresponds to the free-fall case, with a power~law coefficient of $q \lesssim$ 1.5.  
For higher values $(q > 1.5)$, \citet[][]{Franco1990} showed that expansion is in  
the so-called "champagne" phase.
In such cases, there is neither velocity damping nor creation of a collected shell   \citep[see also][]{Shu2002,Whalen2006}. 
In contrast, the shell and low expansion velocity observed for the central UC\hii region \citep{Fuente2010,pilleri2012,Pilleri2013,Pilleri2014} indicate that $q=1.5$ is 
an upper limit. 
The ionisation front should then stay behind the infall wave, $R_\ion{H}{ii}< R_\text{Infall}$, with the expansion starting at supersonic velocities, 
classically $\sim$10~km$\,$s$^{-1}$, but then quickly damped {to values around the sound speed in the neutral molecular medium}.
As a matter of fact, the expansion of the central UC\hii slowed down
to $\sim$1~km$\,$s$^{-1}$ at $\sim$0.08/0.09~pc \citep[][and references therein]{Pilleri2014}, so far behind the infall wave situated around 0.3 pc { (see Table~\ref{table_Envchar})}.
More generally, in the case of the four \hii regions of Mon~R2, 
{ even if the initial ionisation extension exceeded the central core size $R_{\rm Str} > r_{\rm c}$  },
ionisation would expand in internal infalling envelopes with shallow ($q<1.5$) density profiles.
The $\rho_\text{1}$ value used in Eq.~\ref{rhoprof} corresponds to the internal envelope, $\rho_\text{int}(\rm1pc)$, 
and is calculated from the one observed in the external envelope, $ \rho_\text{env}(\rm1pc)$,  by the relation 
 \begin{equation}
 \label{2rho1}
\rho_\text{int}(\rm1pc)  = \rho_\text{env}(\rm1pc)\times R_\text{Infall}^{( q_\text{in} - q_\text{out}) }  \end{equation}
where $\rho_\text{env}(\rm1pc)$, $R_\text{Infall}$, $q_\text{in}$, and $q_\text{out}$ are given in Table~\ref{table_Envchar}.

For the central core, 
we adopted a size of 0.005~pc ($\sim$1000~au), which is meant to represent the size where the spherical symmetry is broken. 
Since 3D effects are beyond the scope of this paper, gas on small scales (<0.005~pc) is represented by an homogeneous and constant density medium. 
This value is between the typical size of Keplerian disks surrounding low-mass pre-main sequence T Tauri stars \citep[$\sim$0.001pc, 200 au,][]{Carmona2014,Maret2014,Harsono2014} 
and {the largest} disks or flattened envelopes observed around intermediate- to high-mass Herbig Ae-Be stars 
 \citep[ 0.01~pc, $200-2000$~au,][]{Chini2006,deGregorio2013,Jeffers2014}. 
 Calculations with various $r_c$ values in the conditions considered here ($ q  \lesssim 1.5$) show that when
$r_\text{c} \ll R_{\rm \ion{H}{ii}}$, the value of $r_\text{c}$  does not have much influence on the expansion time of the ionised bubble.

For objects studied here, the initial size of the Str\"omgren sphere, $R_{\rm Str}$, is always smaller than the core radius, $r_\text{c}$.
 Therefore the expansion is decomposed in two phases: the expansion in an homogenous medium of constant density, $\rho_\text{c}$ from   $R_{\rm Str}$ up to $r_\text{c}$  
 and the expansion in the decreasing density part of the envelope between $r_\text{c}$ and $R_{\rm \ion{H}{ii}}$. 
The expansion time of the first phase in the central core, $t_{\rm c}$, is given by the Spitzer formula (Eq.~\ref{exptime}), and 
the second {phase in the decreasing density envelope}, $t_{\rm d}$, is given by Eq.~\ref{exptimeDecrEnv}.
The total expansion time $ t_\text{exp} =  t_{\rm c} + t_{\rm d}$, is given in Col.~5 of Table~3. 
It is consistent with the value obtained for an expansion within a homogenous-density medium with an equivalent mass
(see Col.~3),  corroborating the constant-density equivalence {hypothesis and the calculation performed in Sect.~\ref{sec:HIIRageCalcCstEnv}}.
Agreement is better for the compact or UC \hii regions, which have $\langle\rho_\text{initial}\rangle$ closer to $ \rho_\text{c}$ than the extended northern region.

The time for an \hii region to expand depends strongly on the density gradient in the inner envelope through the  
density extrapolated for the 0.005 pc core, $\rho_{\rm c} $, which determines the Str\"omgren radius, $R_{\rm Str}$. 
The gradient index, $q_{\rm in}$, itself gives the transit time in the envelope, $t_{\rm d}$.
All measures are uncertain, especially because the \hii region expansion occurred in an envelope whose density may be higher than the one presently observed.
If ionisation expansion took place in a primordial envelope with a density slope typical of free-fall, 
$q_{\rm in} = 1.5$,  then the calculations would give a minimum age of $10^5$ yr for all three compact and UC \hii regions.

\subsection{Simulations of  \hii regions development with gravity}\label{sec:simus}

We used dedicated simulations to assess  the impact of a decreasing density and the effect of gravity on our estimates of the ionisation expansion time.
We employed the \text{HERACLES}\footnote {Available at http://irfu.cea.fr/Projets/Site\_heracles } code \citep[e.g.][]{Gonzalez2007}, which is 
a three-dimensional (3D) numerical code solving the equations of hydrodynamics and which 
has been coupled to ionisation as described in \citet{Tremblin2012b, Tremblin2012a}. 
The mesh is one dimensional in spherical coordinates with a radius between 0 and  2~pc and a resolution of 2$\times 10^{-3}$~pc, corresponding to 10 000 cells. 
The boundary conditions are reflexive at the inner radius and allow free flow at the outer radius. 
The adiabatic index $\gamma$ defined as $P \propto \rho^\gamma$ is set to 1.01, 
allowing us to treat the hot ionised gas and the cold neutral gas as two different isothermal phases.

The effects considered here are 1) the ionisation expansion and 2) the gravity of the central object. 
The envelope is considered at rest at the beginning of the simulation, but when gravity is taken into account 
the infalling part of the envelope has a velocity close to the expected values.

Our models consider the gravitational effects of the ionising star but not that of the whole cloud, which could collapse under its own gravity. 
Such gas infall is important but modelling it is beyond the scope of this paper.
It requires complex 3D turbulent simulations with self-gravity such as those developed in Geen et al. (in prep.).
Their main conclusion is that global cloud infall can stall the expansion of \hii regions
but do so less efficiently than 1D arguments. Indeed, infall gas should generally be inhomogeneous and clumpy, 
and the \hii region should expand in directions where fewer dense clumps are present. 

Realistic winds from early B type stars have been checked to make certain they do not quantitatively modify the ionisation expansion time. 
A wind cavity in the ionised gaz would indeed only marginally increase the electronic density. 

Important leaking and outflows of ionised matter would have a noticeable impact and is observed neither in the temperature map  
as seen by \citet{Anderson2015} for RCW120 nor in the hydrogen RRL or the radio data. 
Disruption of the molecular cloud thus cannot be important in the UC\hii region. 
Moreover, the ionising photon flux we used comes from the HII region cavity. 
Any leaking of ionised matter would decrease the amount of ionised gaz in the cavity and thus increase the expansion time. 

Complex structure and motions of the ionised gas have been  reported by \cite{Jaffe2003} and modelled by \cite{Zhu2005, Zhu2008}. 
They argue for tangential motions of the ionised gas along the surface of the HII region, which cannot suppress the thermal pressure at the origin of the expansion evolution. 
Thermal turbulence of ionised gas, the prime driver of expansion, is taken into account in simulations. 
Recent works show that introducing additional turbulence does not affect the HII region expansion
much because it would only increase the expansion time weakly \citep[see][]{arthur2011,Tremblin2012b, Tremblin2014}. 

Taking all these complexities and secondary effects into account is beyond the scope of the present paper since we merely tried to reproduce an overall behaviour, using the mean density at the origin.  Detailed studies would require to take departure from spherical symmetry on small scales into account.   

Simulations are based on 1) the density profiles derived from column density and 2) UV ionizing fluxes deduced from radio data. 
Temperature or electronic density are not used in the initial conditions so any changes of their values would not affect the simulation results.
 
For comparison, we used the profiles presented in Sect. \ref{sec:HIIRageCalcDecrEnv} to perform simulations of \hii regions expanding within decreasing density envelopes 
without taking the gravity into account.
The results  given in Col. 6 of Table~\ref{table_ionitimes} agree within 15$\%$ with analytical calculations given in Col.~5. 
The remaining differences arise from assumptions made in the analytical solutions, which neglect the inertia of the shell and strong external pressure. 
These expansions are illustrated in Fig.~\ref{fig:simuCentralExpansion}-\ref{fig:simuNorthernExpansion} by black lines.

We hereafter investigate the effect of gravity.
Section~\ref{sec:simusAvecG} shows that extrapolating density profiles as done for analytical calculations of Sect.~\ref{sec:HIIRageCalcDecrEnv} would lead to quenching or recollapse
for the central, western, and northern \hii regions.
 For the eastern region, the extrapolated  low central density is not high enough for gravity to impede expansion, so expansion time remains similar.
In Sect.~\ref{sec:HIIsimuCstEnvWithG}, we calculate the maximum constant-density medium inside which the \hii regions could develop.
{Unrealistic envelope models with constant density would induce re-collapse even on a large scale, but they nevertheless provide useful upper limits to the ionisation expansion time}.
The case of the less dense {northern and eastern} \hii regions is discussed {again} in Sect.~\ref{sec:HIIsimuNorthEnvWithG}, 
{and we confirm that once the expansion starts, the gravity influence is negligible, at least in decreasing-density envelopes}.
In Sect.~\ref{sec:HIIsimuConclusion}, we summarise the results {we obtained from simulations} 
and give estimates of the ionisation expansion time for the four \hii regions in Mon R2.

\subsubsection{Effect of gravity: quenching or recollapse}\label{sec:simusAvecG}

Introducing gravity due to the central object in the simulations results in the quenching {\citep[also called "choke off" by][]{Walmsley1995}} of 
{three} \hii regions  discussed here, and they expand in high-density media: central, western, and northern regions. 
They cannot develop and are {\it \emph{quenched}} or {\it \emph{gravitationally trapped}} \citep[][]{Keto2007}, even with the introduction of a realistic wind support arising from early B-type stars. 

The eastern \hii region is not quenched and can develop in the observed inner envelope, which has a weak density gradient. 
The values of  $\langle\rho_\text{initial}\rangle$ and even $\rho_\text{c}$ are lower than the maximum constant density $\langle\rho_\text{initial}\rangle_\text{Max}$ 
(see Table.~\ref{table_ionitimes}), explaining why quenching is not occurring here. 
In this case of a low-density envelope, the expansion time obtained in simulation without gravity (23 kyr, see Col.6 of Table~\ref{table_ionitimes})
is identical when including gravity in simulation  (24 kyr, see Col.9 of Table~\ref{table_ionitimes}).

To allow expansion of {the three} \hii regions {quenched} in simulations, one needs to decrease the central density of the profile, $\rho_c$.
If we keep extrapolating the density profile observed for the outer envelope (see $\rho_1$ in Col.10 of Table~\ref{table_Envchar}),
the only way is to increase the constant density core radius, $r_c$. 
This determines the maximum central density value, ${\rho_c}^\text{Max}$, and the corresponding minimum core radius, ${r_c}^\text{min}$, 
allowing an \hii region  to develop up to the observed size.
For the UC (central) and the compact (western) 
\hii regions, ${r_c}^\text{min}$  is larger than $R_{\rm \ion{H}{ii}}$,  corresponding to an 
expansion in a homogenous medium of density ${\rho_c}^\text{Max}$. 
The highest density observed in the envelope at a radius corresponding to the \hii region size 
is thus greater than the maximum density allowed by this scenario, 
${\rho_c}^\text{Max}  < \rho(R_{\rm \ion{H}{ii}})$, and the radii of constant density are larger than the actual  \hii region sizes, $r_c > R_{\rm \ion{H}{ii}}$.
Even if these assumptions are unrealistic, {they are useful to determine upper limits.}
Another way to reduce the central density, as shown in Eq.~\ref{rhoprof}, is to keep the same central core radius but decrease the density of the entire profile  through the ${\rho_1}$ value. 
This configuration {implicitly} assumes a mean envelope density at the initiation of the \hii region expansion that is lower than expected. 

Figure~\ref{fig:simuCentralExpansion} shows four simulations for the central UC \hii region. 
The expansion in a decreasing-density envelope with no gravity (black lines) is compared to two simulations including gravity and occurring in constant-density envelopes. 
For a constant low-density envelope (cyan curve) expansion can reach the observed \hii region size.
For higher constant density (red dashed curve), the re-collapse occurs {earlier and closer, preventing} the expansion up to its present size. 
Increasing the density gradient of the inner envelope from that presently observed ($q_\text{in}$=0.85, continuous black line)
to that expected for free-falling gas ($q_\text{in}$=1.5, dotted black line) increases the extrapolated density and thus
the expansion time.

Figure~\ref{fig:simuNorthernExpansion} gives the expansion behaviour of five simulations for the northern \hii region. 
As in Fig.~\ref{fig:simuCentralExpansion} and with the same colours, three simulations describe 
expansion within a decreasing density envelope without gravity (black line) and expansion in constant high- and low-density envelopes without gravity (dashed red and continuous cyan curves, respectively).
Two additional simulations in a decreasing density envelope with gravity are displayed by a blue line for the nominal average initial density and a dashed blue line for a lower density.

\begin{figure}[htbp]
\includegraphics[height=.72\hsize]{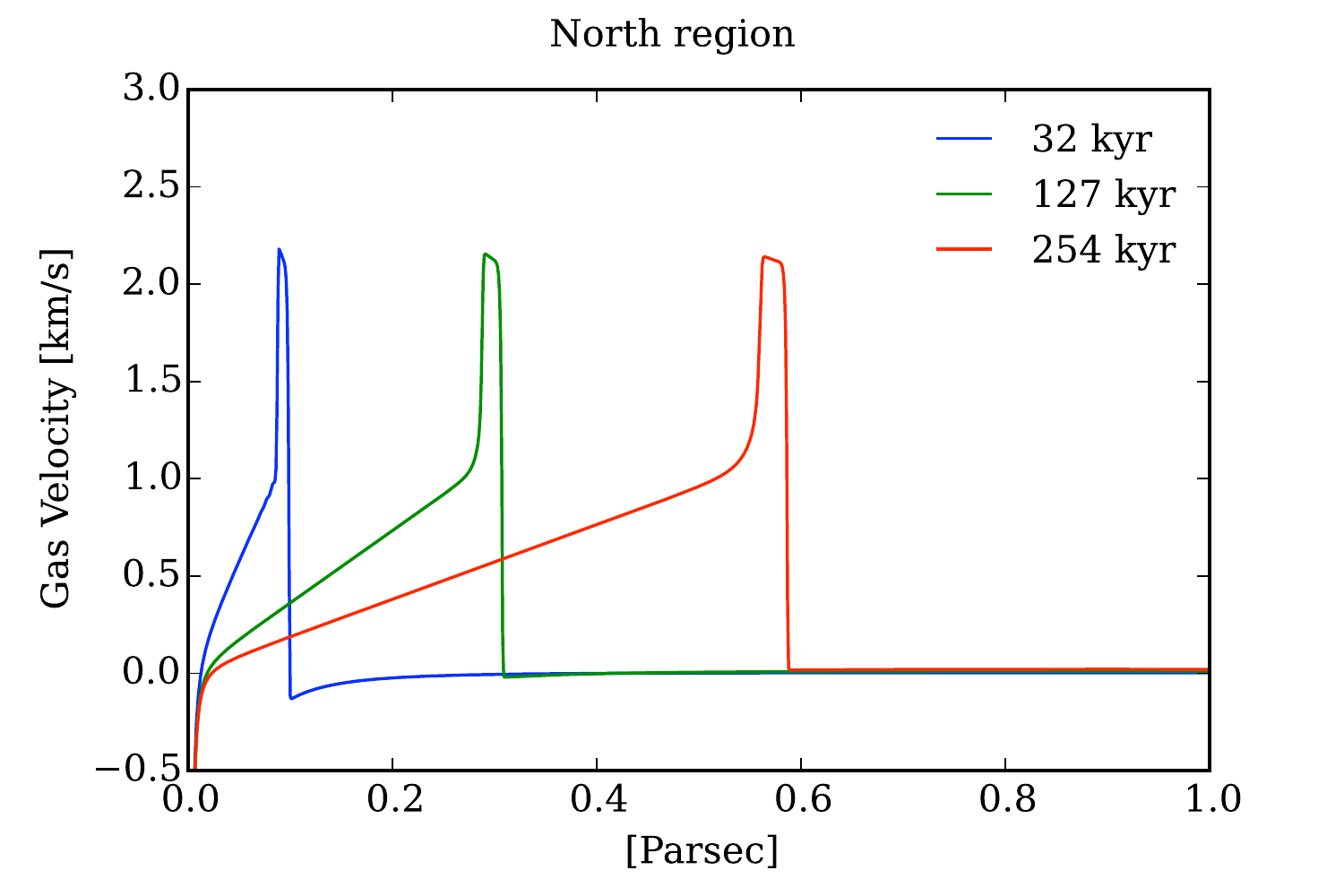}
\hfill 
\vspace{-.8cm}
\caption{ Velocity field from numerical simulations for the northern \hii region expansion at three different times. 
Simulation is made with gravity and in an envelope of decreasing density, which is extrapolated from the observed one up to a constant density within  ${r_c}$ = 0.005 pc.
\label{fig:simuNorthernVelocityField}}
\vspace{-.3cm}
\end{figure}

\begin{figure}[htbp]
\includegraphics[height=1.\hsize]{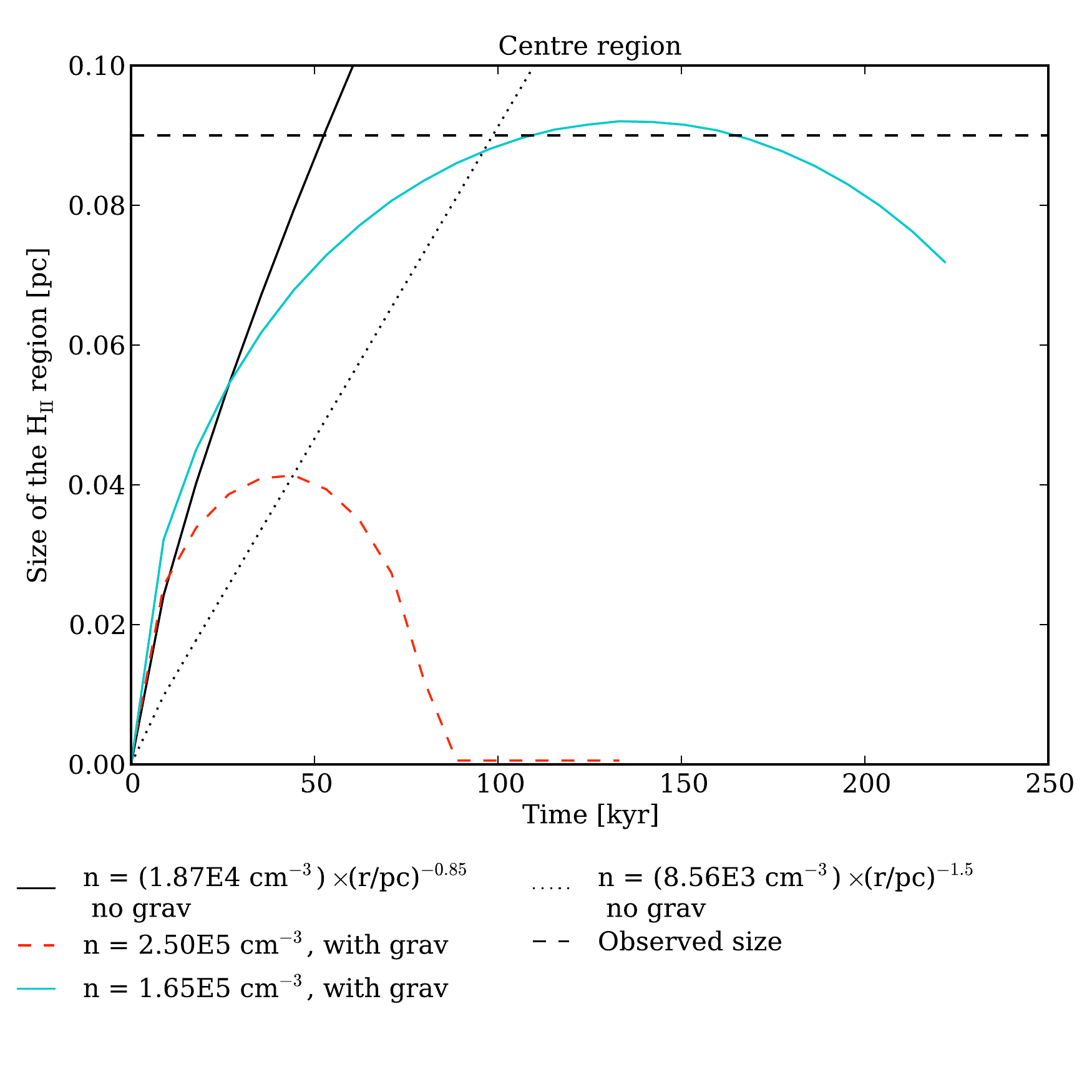}
\hfill 
\vspace{-1.3cm}
\caption{ Numerical simulations for the central UC \hii region  expansion. 
In black: without gravity and in an envelope of decreasing density, which is extrapolated from the observed one up to a constant density within  ${r_c}$ = 0.005 pc.
{In dotted black: without gravity in an envelope of decreasing density with a gradient typical of infall, $\rho \propto r^{-1.5}$.}
In cyan: with gravity and in an envelope with a constant density of $1.65\times10^5$~cm$^{-3}$, just allowing the observed size to be reached (black dashed line).
In dashed red: with gravity and in an envelope with a constant density of $2.5\times10^5$~cm$^{-3}$, collapses again before reaching the observed size. 
\label{fig:simuCentralExpansion}}
\vspace{-.3cm}
\end{figure}

{
Figure~\ref{fig:simuNorthernVelocityField} shows the velocity field of a simulation for the northern region obtained at three different times. 
The simulation includes the gravity of the central ionising object in a density decreasing envelope with a central core of 0.05 pc and a density of 110 cm$^{-3}$ at 1 pc.
This envelope has a mean density in agreement with the observed characteristics (see Table~\ref{table_Envchar}).
It corresponds to the blue continuous curve in Fig.~\ref{fig:simuNorthernExpansion}.
The gravity influence is only noticeable at small scales where gas velocity is negative corresponding to infall, proeminent mainly in the ionised bubble.
The effect is less pronounced outside the shell and get weaker when the \hii region size increases.
}

\subsubsection{Case of constant-density envelopes}\label{sec:HIIsimuCstEnvWithG}  

We simulated the expansion in a constant density envelope, whose value {could be compared to} the initial average density of the envelope material swept up by the expansion.
The latter is  estimated through Eq.~\ref{rhoinit} (see Col.~2 of Table~\ref{table_ionitimes}).
This approach is validated in the case of the northern region (see Sect.~\ref{sec:HIIsimuNorthEnvWithG}) 
and is shown to be a good approximation for {all}  \hii regions since, for analytical calculations without gravity,
the expansion times given in Cols. 3 and 5 of Table~\ref{table_ionitimes} are similar. 

The central UC \hii region and the western compact  \hii  region are quenched by gravity when densities approach the initial average density extrapolated from the observed envelope. 
For these two regions, we thus decreased the envelope average density until their \hii regions develop and reach their actual size. 
The associated expansion times are then upper limits measured for the densest, homogeneous, and constant density envelopes.
Any density above this value would result in quenching or re-collapse before reaching the observed size. 
Any configuration that allows expansion up to the observed size leads to a shorter expansion time.
The corresponding {maximum values of the initial homogeneous constant} density and expansion time are given in Cols.~7-8 of Table~\ref{table_ionitimes}.

Figure~\ref{fig:simuCentralExpansion} shows the effect of gravity  on the expansion for the central UC \hii region in a constant low-density envelope (cyan curve) 
compared to the expansion in a decreasing density envelope constrained by observational data, but without gravity (black line). 
For higher density (red dashed curve), the re-collapse occurs, preventing the expansion to reach the observed size. 
The same re-collapse behaviour is observed for the western compact \hii region.
At the beginning of the expansion, {corresponding to small scales}, gravity can have a strong influence.
In the case of a constant-density envelope, the collected mass noticeably increases when size grows. 
The re-collapse of the shell is thus favoured in a constant-density envelope, but it occurs less easily  
in a {more realistic, decreasing-density} envelope or at larger size.
At the beginning of the expansion, the expansion is faster, even with gravity, in the constant-density cases. 
This arises from the fact that the central density here is much lower than in the case without gravity of extrapolated envelopes with decreasing density. 
When the collected mass becomes important, the expansion is strongly slowed down by gravity and even stops. 
Re-collapse then occurs, at large distance, but a constant density is unrealistic for such large sizes.

\begin{figure}[htbp]
\includegraphics[height=1.\hsize]{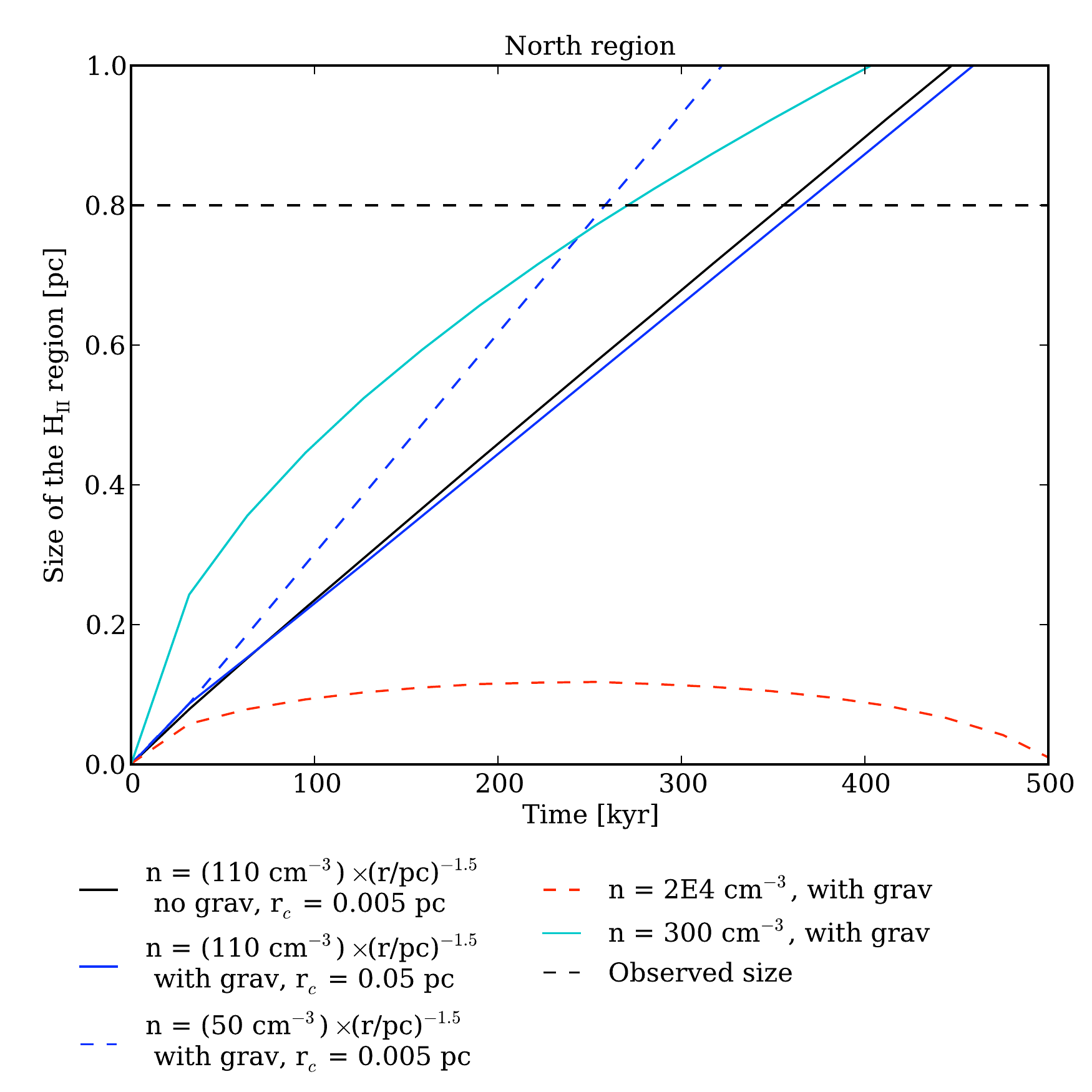}
\hfill 
\vspace{-.8cm}
\caption{ Numerical simulations for the northern \hii region. 
Simulations with and without gravity are in colours and black, respectively.
Continuous lines correspond to simulations with similar average initial density, $\langle\rho_\text{initial}\rangle = 300$~cm$^{-3}$.
The envelope has a decreasing  power-law density with the characteristics of the observed one and extrapolated to a constant core of radius 0.005 pc (black) and 0.05 pc (dark blue). 
For the blue dashed line, the envelope has a power~law decreasing density with an index corresponding to the observed gradient but  
with a $50$~cm$^{-3}$ density  at 1 pc extrapolated to a constant core at 0.005 pc. 
For the cyan and red dashed curve, the envelope has a constant density of $300$~cm$^{-3}$ (cyan),  corresponding to the average initial density,
or a higher $2\times10^4$~cm$^{-3}$ density (red dashed),  leading to re-collapse.
\label{fig:simuNorthernExpansion}}
\vspace{-.5cm}
\end{figure}

The northern \hii region could expand up to its observed size in an envelope with a maximum constant density of $5\times10^3$~cm$^{-3}$.
This density is much higher than the mean value extrapolated from observations, and 
the observed size would be reached in a very long and unlikely time of at least $3\times10^6$  yr. 
At this large distance, re-collapse will not occur. 
More realistically, the northern \hii region should have expanded in an envelope with constant density of 300~cm$^{-3}$ (cyan curve in Fig.\ref{fig:simuNorthernExpansion}), 
equal to the initial average density deduced from the envelope presently observed. 
This gives an expansion time of $\sim$300$\times10^3$ yr. 
As for the central  UC\hii region, the constant-density envelope case has a low density at the centre. which favours the expansion at the beginning (cyan curve). 
But on large scales the expansion times of all models calculated for the nominal constant  average density envelope  
converge to similar values. The same situation applies to the eastern \hii region. 
Better estimates of the expansion time can be obtained for the northern and eastern \hii regions, as shown in the Sect.~\ref{sec:HIIsimuNorthEnvWithG}.

\subsubsection{Case of decreasing-density envelopes}\label{sec:HIIsimuNorthEnvWithG}

The two options already mentioned in Sect.~\ref{sec:simusAvecG} for reducing the central density are suitable for the northern \hii region.
The first option is to increase the radius of the constant density core, $r_c$.
The lower density and larger size of the northern region allows expansion with gravity without quenching  up to the  \hii region size. 
Nevertheless, the required size of the central core, $r_c \sim$ 0.05 pc (10000~au), is then much larger than those inside  which we expect 
{a departure from the spherical symmetry ($\sim$0.001~pc/200~au - 0.005~pc/1000~au)}.
As in the case of a constant-density envelope, the initial conditions and the central structure of the  \hii region has less of an impact 
on the analysis of the old and more developed northern \hii region since the expansion has already taken place for a longer time. 
This explains why the expansion times are not very different with and without gravity (blue and black continuous lines in Fig.\ref{fig:simuNorthernExpansion}).
The highest central density allowing an \hii region expansion up to the observed size defines the maximum time of $370\times10^3$ yr (see Col.~9 of Table~\ref{table_ionitimes}). 

The second option is to lower the density of the overall structure, and in the northern \hii region case, a factor of two is necessary. 
The density thus decreased by 2, the expansion time is noticeably reduced  as shown in Fig~.\ref{fig:simuNorthernExpansion} by the blue dashed line. 
From Eq.~\ref{exptimeDecrEnv} ($t_{\rm exp} \propto  {{R_{\rm Str}}^{-3/4}} $) 
and Eq.~\ref{eq_rstromgren0} ($ R_{\rm Str} \propto \rho^{-2/3} $), 
the expansion time relates to density through its square root: $t_{exp}\propto\rho^{1/2}$.  
This agrees with the time approximately divided by $\sqrt{2}$ between the two cases, 
$\rho_1$ and $\rho_1/2$ (blue and blue dashed lines in Fig~.\ref{fig:simuNorthernExpansion}).

In fact the average initial density, $\langle\rho_\text{initial}\rangle$, is identical for the three cases best representing the northern region. 
Case 1 is the first option mentioned above in this section with gravity and an enlarged central core (blue line), 
Case 2 is the simulation without gravity (black line), 
and Case 3 is for a constant density with gravity (cyan line). 
Regardless of whether the gravity is included in simulations, these three lines lead to similar expansion times. 
It indicates that the average initial density in which the \hii region has expanded is one of the important parameters for estimating expansion time. 
Once {quenching by} gravity has been overruled, the expansion is no longer strongly affected by gravity.
In the details, however, the configuration with constant density and gravity underestimates the expansion time mainly in the central parts where the density should be much higher than what is estimated. 

For the northern \hii region, expansion simulated in a decreasing envelope without gravity gives a minimum expansion time of 355 kyr. 
Expansion  simulated with gravity in a decreasing-density envelope with the highest possible central density gives a maximum time of 370 kyr.
These values define a very narrow interval for the expansion time, if extrapolation of the observed envelope towards the interior is valid. 
Taking the observational uncertainties into account, including those concerning the small scales structure, 
we conclude that the expansion time of the northern \hii region should be 350+/-50~kyr.

Even more strikingly, for the eastern  \hii region, we do not need to increase the default central core radius (0.005 pc), 
and the time expansion calculated with gravity (24 kyr) is almost identical to the time obtained without gravity (23 kyr).
This shows again that gravity could delay the start of expansion, but has very small influence on expansion time. 
This effect is more pronounced in a low-density medium.

\subsubsection{Ionisation age and expansion time}\label{sec:HIIsimuConclusion}

Tracing the expansion of \hii regions with gravity and in a decreasing-density envelope extrapolated from the one observed at large scales 
would require complex 2D or even 3D configurations, at least on small scales. 
On large scales, in a decreasing-density envelope with a power~law index $q \gtrsim 1$, once gravity has been overcome by turbulence and expansion occurs, 
the gravitational pressure decreases faster than the turbulent pressure, and its influence becomes negligible 
(\citealt{tremblin2014HRDSlarson}, section~5.1, Eq.~12).  
Accretion within a rotating envelope can create a lower density zone where infall is driven from the envelope onto a disk or  torus 
(see \citealt[][Fig.10]{Tobin2008}; \citealt[][Fig.1]{Hosokawa2010}; \citealt[][Fig.5]{Ohashi2014}). 
The decreased density near the ionising objet reduces the influence of gravity. 
It should result in a reduction  of the expansion time and decrease in re-collapse probability. 
Outflow lobes also modify the density structure and the geometry of the envelope (outflow cavity) and thus influence the expansion time. 
Infalling gas from the envelope itself locally increases the density and reduces the expansion speed or impedes it in a disk, torus, or funnel. 
All these effects, along with sporadic accretion  \citep{Peters2010acc,Duarte-Cabral2013,DePree2014}, increase the expansion time uncertainty. 
A definitive measurement of the ionisation age would thus require a detailed modelling based on better constraints of physical parameters, at least on small scales.
Unfortunately, the observational constraints of  high-mass protostellar envelope on small scales (< 0.05 pc) are still {largely} missing. 
This 3D small-scale asymmetric geometry could also shorten the quenching time, which needs to be added to the expansion time to determine the age of \hii regions and their ionising stars. 
On the other hand, confinement by external pressure could also occur at the end of expansion and thus lengthen the ionisation age. 
Expansion times depend on the initial conditions on small scales 
and especially ionisation age if the quenching and confinement times are taken into account.
However, the order of magnitude of the expansion times determined from simulations seems to converge (see Table~\ref{table_ionitimes}). 

The simulations performed with or without gravity for the eastern and northern \hii regions expanding in low-density envelopes 
provide expansion times of $25 \pm 5 \times 10^3$~yr and $350 \pm 50 \times 10^3$~yr. 
For the central UC and western compact \hii regions, we estimated upper and lower limits for the expansion time. 
Lower limits correspond to simulations in a decreasing-density envelope without gravity (Table~\ref{table_ionitimes}, Col.~6).
Upper limits are given by simulations with gravity in an envelope with the highest constant density (Table~\ref{table_ionitimes}, Col.~8)
avoiding quenching or re-collapse and allowing the observed size to be reached.
The ranges of time values obtained, $90 - 150 \pm 50 \times 10^3$~yr,
are consistent with a mean expansion time of $1   \times 10^5$~yr. 
The expansion time of the central UC \hii region, $9  \times 10^4$ yr, agrees with the duration of the warm phase in dedicated chemical models \citep[$10^4$ - a few $10^5$~yr,][]{Trevino2014}. 
{The chemical model requires a chemistry out of equilibrium and imply a warm phase duration of less than $10^6$~yr.}

Despite the dependence on unknown small scale geometry, 
expansion times adopted for the dense western compact and central UC \hii
regions are equal to within a factor two, converging to $\sim$1-1.5 $\times 10^5$ yr (Table~\ref{table_ionitimes} Col.~10).
They agree with the statistical estimate of the age of C and UC \hii regions \citep[$\sim$$10^5$ yr,][]{Wood1989} 
and are ten times longer than the rough dynamical time ($\sim10^4$ yr) estimated from the initial ionised gas velocity. 
As already stressed by \citet{Urquhart2013}, the order of magnitude of these ages, $3-6 \times 10^5$~yr, 
 has recently been confirmed by new statistical lifetimes \citep[$\sim$3$\times 10^5$~yr,][]{Mottram2011} 
and supported by a Galactic population synthesis analysis \citep[$2-4 \times 10^5$~yr,][]{Davies2011}. 
The so-called lifetime problem \citep{Wood1989,Churchwell2002,peters2010} 
is simply solved here by taking an expansion velocity equal to the sound speed in ionised gas ($c_s$) only at the beginning.  
Indeed, in a slowly decreasing-density envelope ($q<1.5$) expansion speed is rapidly damped 
and even stopped if confinement {by external turbulent pressure}  occurs \citep{Raga2012}.

\subsection{Suitable approximations for expansion time estimation }\label{sec:gravDensApprox}

As already mentioned in the previous section, gravity does not itself strongly affect the expansion time, but
a delay of expansion can occur at the beginning.
Similarly, \citet[][]{Keto2007} differentiates the dynamic behaviour of HC and UC \hii regions, which are gravitationally dominated, 
from that of larger compact and extended \hii regions, which are thermally dominated.
As a matter of fact, 
in the low-density conditions of the northern and eastern regions, simulations with or without gravity give similar expansion times  (see Table~\ref{table_ionitimes}, Cols.~6 and 9).
In the high-density condition of the central and western regions, the upper limit of expansion times calculated with gravity (Table~\ref{table_ionitimes} Col.~8),
even with a rough constant density estimation, are only twice longer than expansion times obtained without gravity (Table~\ref{table_ionitimes} Col.~6).

Even if the average initial densities ($\langle\rho_\text{initial}\rangle$, Table~\ref{table_ionitimes} Col.~2) 
are only roughly estimated, expansion times of the various analytical calculations and simulations agree (Table~\ref{table_ionitimes}, Cols.~3, 5, and 6).
The agreement is better for the small (C and UC) \hii regions, for which the average initial density is more representative of the overall {past density} conditions.
Analytical calculations without gravity using an extrapolation of the observed density envelope give a good time estimate at least for \hii regions of small sizes. 
For the more extended northern \hii region, the evaluation made of the average envelope density on small scales is 
more questionable since the remaining envelope has much lower density than the one at the beginning of expansion.
Calculations using more realistically extrapolated density profiles should be preferred, or
other evaluation using the pressure equilibrium and third Larson law  \citep[][]{tremblin2014HRDSlarson} could be made. 

We conclude that:

- Gravity induces quenching, which adds a delay before the \hii region expands, but when it expands, the influence of gravity becomes marginal. 

- Envelope density profiles could be used to estimate a {minimum} expansion time from analytical calculations without gravity,
 and if density is not too high, it gives a realistic value.
 
- The mean initial density is a good estimator of the matter density surrounding the {massive} protostar in early stages and is later collected in the shell by the \hii region expansion. 
It could be used to estimate expansion times if density profiles are not well defined. In any case, it is better suited 
{\bf to describe the intial density condition of ionisation expansion} than the surrounding cloud density 
characterising the external medium.

All realistic time expansion estimates done in this paper converge around $\sim$$1 \times 10^5$~yr for the two dense (UC central and compact western) \hii regions,  
 $\sim$$ 2.5 \times 10^4$~yr for the  compact eastern \hii region, and 
$\sim$3.5$\times 10^5$~yr for the extended northern \hii region.

\section{History of the formation of the B-type stars powering \hii regions in Mon R2}\label{sec:HIIRage}
In Sect.~\ref{sec:globalinfall}, we discuss the global 
infall of the Mon~R2 cloud. In Sect.~\ref{sec:protinfall}, we analyse the two-slope profiles 
in the framework of the inside-out collapse of envelopes and estimate protostellar infall ages. 
In Sect.~\ref{sec:agecomp}, we compare the infall and ionisation ages and propose a full history of massive star formation in the Mon R2 cloud.
 
 \subsection{Cloud {global}  infall}\label{sec:globalinfall}
The density structure and infall kinematics discussed here are probably dominated by the most massive object clearly identified in Fig.~\ref{fig:hotRcoldTmaps}b.
The envelope profile defined above (see Sect.~\ref{sec:HIIRdensstruct}) and their corresponding values given in Table~\ref{table_Envchar} 
have been used in Eq.~\ref{mass2} to estimate a mass of $\sim$3500 M$\odot$ in a 2.5 pc radius for the central  \uchii region. 
This region exhibits a steeper density gradient in its outer envelope, with $q_\text{out}\simeq-2.5$, 
{also illustrated by} the cloud structure studied with probability density function (\citet{Schneider2015}, Rayner et al. in prep.).
It suggests compression by external forces, as shown in numerical simulations of \cite{Hennebelle2003} and outlined in Appendix A. 
This compressive process could be associated with stellar winds and ionisation shocks of nearby stars
such as those observed in the M16 massive YSO (young stellar object) by \citet{Tremblin2014} 
or with 
{globally} infalling gas driven by the dynamic formation of Mon R2 cloud through force fall.
Such infalling motions have in fact been observed in CO by \cite{loren1977}, and 
clumps with similar masses generally exhibit active 
{global} infall.
This is the case of SDC335, which has a mass of $\sim$5500 M$\odot$ in 2.5 pc  \citep{Peretto2013} and contains a protostellar object of similar type B1 \citep{Avison2015},
as well as of DR21, Clump-14 (DR21-south, $\sim$4900 M$\odot$), and Clump-16 (DR21-north, $\sim$3350 M$\odot$) \citep{Schneider2010},
all of which show supersonic infall (V=0.5-0.7 km.s$^{-1}$).
Moreover, the hourglass morphology of the magnetic field is thought to be a signature of 
{global} infall \citep{Carpenter2008} ,
and \citet{Koch2014} suggest that Mon R2 is a super critical, 
quickly collapsing cloud.
The expected infall velocity gradient, {as} observed in SDC13 \citep{Peretto2014}, seems to be a crucial ingredient 
for generating the filament crossing \citep{Dobashi2014} characteristic of all the regions mentioned here and commonly observed in many places.
 
 \subsection{Time elapsed since the beginning of the protostellar infall}\label{sec:protinfall}
 The outer envelope of the two compact eastern and western \hii regions display a column density index of $p_\text{out} \simeq -1\pm0.2$, 
 which corresponds to a $\rho(r)\propto r^{-2}$ density law.
It recalls the density distribution of the singular isothermal sphere \citep[SIS,][]{Shu1977}  and that of the cloud structures quasi-statically forming clumps in numerical simulations with or without turbulence and magnetic field  \citep[e.g.][]{LiShu1996}.
A similar density distribution is in fact obtained for clumps forming from subsonic infall \citep[][]{Dalba2012} or supersonic flows \citep[][]{Gong2009}.
This is also reminiscent of the density profiles found for low-mass and high-mass protostellar envelopes in early stages \citep{Tsitali2013}, which has been
known for a long time \citep[e.g.][]{Motte2001,Beuther2002}.
Numerous pre-main sequence phase YSOs are found in the Mon~R2 cloud, suggesting that a group or cluster of protostars has most probably formed within the same area 
of these four separate envelopes of $\sim$2.5-3~pc radius each. 

The transition radius between the inner and outer envelopes, $R_\text{Infall}$, could be used to locate the front of the infall expansion wave associated with the protostellar collapse. 
Indeed, in the case of an inside-out collapse developing into a static envelope with a SIS density structure, the envelope matter is free-falling as soon as the infall front wave travelling outwards reaches its location \citep[e.g.][]{Shu1977}. 
Interestingly, the density profile measured for the inner envelope surrounding \hii regions is close to that of free-falling material, $\rho(r)\propto r^{-1.5}$ (see Table~\ref{table_Envchar}). 
The transition radii would correspond to late stages of the protostellar collapse since they are rather large compared to typical protostellar envelopes: $R_\text{Infall}\sim0.3$, 0.5, 0.9, 2~pc versus $R_\text{prot} \sim 0.03-0.1$~pc \citep[e.g.][]{Motte2001}.
This free-falling envelope material may not be able to reach the stars but may pile up at the periphery of the \hii regions, located at $R_{\rm \ion{H}{ii}}$. 

 We assumed below that the infall front wave initiated at the time of the protostellar embryo formation and propagating outwards from the centre can be approximated by $R_\text{Infall}$. 
 The latter was measured to be 0.3, 0.5, and 0.9~pc for the central, western, and eastern ultra-compact and compact \hii region envelopes (see Table~\ref{table_Envchar}). 
We used an infall front-wave velocity equal to the isothermal sound speed, $a_s = c_s =  \sqrt{ k\,T/\mu\,m_\text{H}}  \simeq 0.2$~km\,s$^{-1}$, calculated for $T = 13.5~$K, which is the mean cloud temperature in Mon~R2. 
We then dated the beginning of the protostellar infall using the following equation:
\begin{equation}
\label{inftime}
t_\text{Infall} = \frac{R_\text{Infall}} {c_s} \simeq 4.9\times10^6\rm~yr \times \frac{R_\text{Infall}}{1\rm\,pc} \times 
 \left( \frac{c_s} {0.2\rm\,km\,s^{-1}} \right)^{-1}. 
\end{equation}
It dates back to a few million years ago for the three compact and ultra-compact \hii regions with $t_\text{Infall}\simeq1.5\times 10^6$~yr for the \uchii 
and $t_\text{Infall}\sim2.5-4.5\times 10^6$~yr for the eastern and western \hii regions.
The more developed northern \hii region has an envelope that could be fully free-falling up to a radius reaching the background density (see Table~\ref{table_Envchar}). 
With a value of $R_\text{Infall}> 2$~pc, our calculation leads to a minimum infall age of $t_\text{Infall}>10^7$~yr.
These ages perhaps {seem too long, but they}
are approximate values since a more complex density, temperature, and kinematic structure is to be expected for a non-isothermal envelope that is already globally collapsing at the time of the creation of the first protostellar embryo. 

Moreover, the static SIS initial conditions are probably far from realistic, particularly for clouds forming high-mass stars, which are known to be dynamical entities \citep{Schneider2010, Peretto2013}.
Some models examined the protostellar collapse in an infalling cloud \citep{Larson1969,Penston1969,Gong2009,Keto2015}. 
The initial 
inflow speed before protostellar collapse indeed modifies 
directly or indirectly the rarefaction wave speed, even in subsonic mode \citep{Dalba2012}.
In the case of supersonic flows, the speed increases by a factor 2 to 3 \citep[][see their section 4.2]{Gong2009}, dividing the calculated infall time by the same factor.
Supersonic inflows driven by the cloud's 
{global} collapse, is probable for the central UC\hii region. 
It is less certain but still plausible for the other \hii regions, which are within the same infalling cloud but excentred from the infall centre. 

Infall times are listed in Col. 6 of Table~\ref{table_Envchar}.
The range of values illustrates systematics and errors to statistical uncertainties.
All in all, it gives the following infall times: 
$\sim$$1.\pm.7\times 10^6$~yr for the central UC \hii region, 
$\sim$$1.5\pm1.\times 10^6$~yr for the western C \hii region, and 
$\sim$$3.\pm1.5\times 10^6$~yr for the eastern C \hii region.
Deducing the ionisation expansion time estimated in Sect.~\ref{sec:ionisationExpansion} 
from this infall time provides a measure of the time a protostar needs to
reach the high-mass regime associated with the emission of ionising UV photons and an eventual 
ionisation delay due to quenching or swelling \citep{Hoare2007}.

\subsection{Age comparison and history of OB star formation in Mon~R2}\label{sec:agecomp}

For the {dense} compact and ultra-compact (western and central) \hii regions,  
infall ages derived in Sect.~\ref{sec:protinfall}: are about ten times greater than the
ionisation expansion time $t_\text{infall}\sim1-2\times10^6$~yr versus $t_\text{ionisation}\sim1-2\times10^5$~yr.
As for the eastern compact \hii region, the infall time is 100 times longer than
the ionisation expansion time: $t_\text{infall}\sim3\times10^6$~yr versus $t_\text{ionisation} \sim25\times10^3$~yr.
The  infall age of the northern extended \hii region is unknown with a lower limit of $t_\text{infall}>14\times10^6$~yr, which is {$\sim$}30 times more than
the ionisation expansion time of $t_\text{ionisation}\sim5\times10^5$~yr.
Even if uncertain, {the order of magnitude of} these values is probably realistic since 
the stellar ages of the B-type star association observed in Mon~R2 is $\sim1-6\times10^6$~yr \citep{HR76,carpenter1997}, 
in agreement with the computed infall ages of the compact and ultra-compact \hii region envelopes.

Subtracting a mean protostellar lifetime of 3 $\times$ $10^5$~yr 
(\citealt[][]{Duarte-Cabral2013}, see also \citealt[][]{Schisano2014} and references therein)
from the infall time measured for the central UC\hii region, 1$\times10^6$~yr suggests that ionisation started ${7}\times10^5$~yr ago.
With an ionisation expansion time of $\sim$$1\times10^5$~yr, the {quenching} {delay} time could be $\sim$${6}\pm5\times10^5$~yr. 
The typical protostellar lifetime adopted here agrees with the one obtained for the cold phase of Mon R2 in chemical models 
($10^5$ - $10^6$~yr  \citealt[][]{Trevino2014}), corresponding to collapse \citep[][]{Esplugues2014}. 
Similar considerations for western (and eastern) C\hii region give a {quenching} {delay} time of $\sim$${9\pm6}\times10^5$~yr (resp. $\sim$${2\pm1}\times10^6$~yr). 
These variations could be interpreted as an increase in the quenching time with increasing spectral type from B0 to B2.5, 
and thus decreasing number of ionising photons $N_\text{Lyc}$.
However, given the large uncertainties of each of these estimates, a constant value around $10^6$~yr cannot be ruled out. 

Besides that, the derived infall ages span an order-of-magnitude difference, possibly indicating progressive star formation in the Mon~R2 cloud.
None of the ionisation fronts {and none of the infall rarefaction wave} arising from the three \hii regions located $\sim$1--2~pc from the \uchii region 
could, however, explain the compression of the \uchii envelope. 
There is thus no clear evidence that star formation in the Mon~R2 cloud could have been triggered by older {local} populations of stars.
Any observed age gradient could only be related to sequential star formation triggered by an external agent.

\section{Conclusions.}\label{sec:conclusion}

We have presented a study of the Mon~R2 molecular cloud based on {dust} column density and temperature maps built from \emph{Herschel}/HOBYS observations and dedicated simuations.
Our main findings can be summarised as follows:\\

1 - The MonR2 molecular cloud is dominated by gas associated with the central UC\hii region. The latter is located at the crossing of three major filaments, while 
three other \hii regions develop in their surroundings. 
The spectral type of ionising stars as estimated from 21cm fluxes are  in very good agreement with those obtained from visual spectroscopy.
They all are early B-type stars.

2 - The size of ionised regions is estimated by the heating of big grains 
(Fig.~\ref{fig:hotRcoldTmaps}a) and small grains excitation through their 70\,\microm\,flux (Fig.~\ref{fig:I70160Profil4center}), 
in the close surrounding of \hii regions, probably within the PDR.
It ranges from $\sim$0.1 pc for compact and UC\hii regions to $\sim$0.8 pc for the most classical extended \hii region.

3 - The \hii regions are surrounded by large and rather dense neutral gas envelopes ($R_\text{env}\sim 2-3$~pc, $\rho_\text{env}^\text{max}\sim 3\times 10^2-5\times10^5$~cm$^{-3}$).
At radii from $\sim$0.1 pc to a few parsecs, the envelope surrounding \hii regions cannot be considered to have a constant density.
Temperature gradients are also observed in the neutral gas envelope surrounding the four regions.

4 - In spite of the quite advanced stage of evolution of the four massive ionising stars 
studied here, the density of the envelope surrounding them keep the imprint of earlier phases of the gravitational collapse. 
The density structure is similar to those expected for individual protostellar objects.
For the first time the power~law slope of the density profile is constrained accurately enough to distinguish between inner layers in free fall ($\rho\propto r^{-w}$ with $w \leqslant 1.5$) 
and external parts that could correspond to an equilibrium SIS configuration ($\rho\propto r^{-2}$). 

5 - We interpret the steep density profile of the central UC\hii region ($\rho\propto r^{-2.5}$) as due to an external pressure certainly associated with the observed 
{global} collapse, 
which can be called a forced~fall.

6 - The transition radius between $\rho\propto r^{-1.5}$ (or $r^{-w}$ with $w \leqslant 1.5$)  and $ r^{-2}$ (or $ r^{-w}$ with $w\geqslant2$) laws should locate the free-fall rarefaction wave. 
Assuming it expands at one to three times the sound speed, we estimate the time since the infall began. 
An infall age around a few million years for these \hii regions is consistent with current high-mass star formation scenarios.

7 - The density profiles obtained here allow us to determine the initial conditions of ionisation and its expansion. 
Dedicated 1D simulations show that the envelope density presently observed on a large scale would induce a complete quenching of the  ionisation expansion. 
A more complex geometry on a small scale and a dynamic scenario are required to explain the  present \hii region development. 

8 - The ionisation expansion time deduced here from analytical calculations and dedicated simulations agree with statistical ages of the corresponding  \hii region. 
The compact and UC\hii region expansion time is $1-2\times10^5$yr.

 Stellar formation in MonR2 seems to be an on-going process that started at least $\sim$$1-6\times10^6$ years ago. 
Large scale feedback, such as the 
{global} infall of the cloud complex, certainly play an important role, but 
no clear evidence of locally triggered star formation has been found.
\\

\begin{acknowledgements}
{P.D. thanks Philippe Laurent for his help in  calculating elliptical integral numerical values.}
This research has made use of TOPCAT (\cite{taylorTOPCAT}, http://www.starlink.ac.uk/topcat/) 
and SAOImage/DS9, developed by the Smithsonian Astrophysical Observatory (http://hea-www.harvard.edu/RD/ds9/).
This work profits from data downloaded from the SIMBAD database, operated at the CDS, and 
the VizieR catalogue access tool (CDS, Strasbourg, France; \citealt{Ochsenbein2000}). \\
SPIRE was developed by a consortium of institutes led by
Cardiff Univ. (UK) and including Univ. Lethbridge (Canada);
NAOC (China); CEA, LAM (France); IFSI, Univ. Padua (Italy);
IAC (Spain); Stockholm Observatory (Sweden); Imperial College
London, RAL, UCL-MSSL, UKATC, Univ. Sussex (UK); Caltech, JPL,
NHSC, Univ. Colorado (USA). This development has been supported
by national funding agencies: CSA (Canada); NAOC (China); CEA,
CNES, CNRS (France); ASI (Italy); MCINN (Spain); SNSB (Sweden);
STFC (UK); and NASA (USA).\\
PACS was developed by a consortium of institutes led by MPE (Germany) and including UVIE (Austria); KU Leuven, CSL, IMEC (Belgium); CEA, LAM (France); MPIA (Germany); INAF-IFSI/OAA/OAP/OAT, LENS, SISSA (Italy); IAC (Spain). This development has been supported by the funding agencies BMVIT (Austria), ESA-PRODEX (Belgium), CEA/CNES (France), DLR (Germany), ASI/INAF (Italy), and CICYT/MCYT (Spain).\\
T.H. was supported by a CEA/Marie-Curie Eurotalents Fellowship. 
Part of this work was supported by the ANR (\emph{Agence Nationale pour la Recherche}) project `PROBeS', number ANR-08-BLAN-0241.

\end{acknowledgements}

\bibliography{./MonR2RegHII.bib} 

\Online

\begin{figure*}[htbp]
\
\includegraphics[height=.99\hsize]{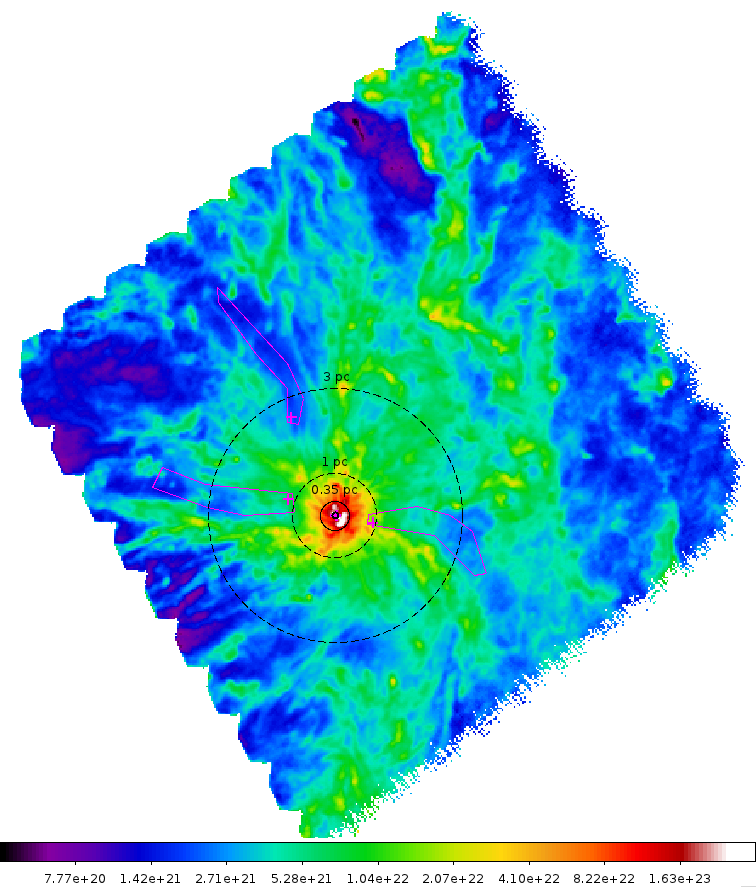}  
\vspace{-.05cm}
\caption{
Column density  map showing the azimuthal selection of areas used to characterise the density profiles of the eastern, western, and northern \hii region envelopes.
The \hii region bubble, the inner, and the outer envelopes of the central UC\hii region are outlined with dashed circles, whose sizes are given in Tables~\ref{table_HIIchar}-\ref{table_Envchar}. 
\label{fig:id_cdens-extraArea4prof}}
\end{figure*}

\begin{figure*}[htbp]
\includegraphics[height=.99\hsize]{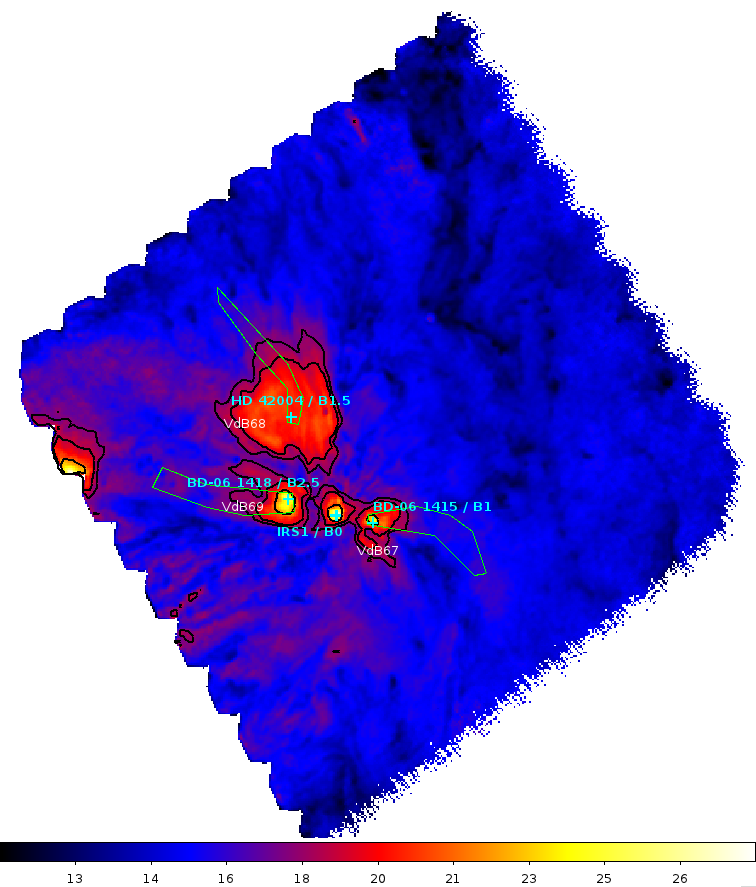} 
\vspace{-.05cm}
\caption{
Hot regions overall contours of profile extraction on temp. map.  
Spectral type are taken from \citet{racine68}. 
\label{fig:id_temp-extraArea4prof}}
\end{figure*}

\clearpage


\begin{appendix}

\clearpage

\section{ \hii region expansion in an envelope of decreasing density with power~law profile }\label{sec:HIIexpansionINdecreasingDensity}

\subsection{Mass and average density from a decreasing power~law density profile }\label{sec:massAverDens}

When considering a decreasing density distribution described by a power~law with an index q, $ \rho \left({r}\right) = \rho_1 {r}^{-q} $, 
the mass included within a radius $R$ is given by 
\begin{equation}  \label{mass1}
M\left( R\right) = 
\int_{0}^{R} \rho_1 {r}^{-q} 4\pi r^2dr =  4\pi \rho_1 \int_{0}^{R}  {r}^{2-q} dr 
.\end{equation}
As far as $ q < 3$, it gives 
\begin{equation}  \label{mass2}
M\left( R \right) =  4\pi \rho_1 \frac{{R}^{3-q}}{3-q} 
,\end{equation}  \\
or in numerical terms
\begin{equation}  \label{mass2num}
M\left( R \right) = 0.76646 \left( \rho_1  \rm\,cm^{-3} \right)  \frac{{ \left( R  \rm\,pc \right)}^{3-q}}{3-q}  M\odot
.\end{equation}  \\
The mean density in the sphere of radius R is then given by
\begin{equation}  \label{mass3rhomean}
\langle\rho\left({R}\right)\rangle =  M\left( R\right) \frac{3}{4\pi  {R}^3} =     \left( \frac{3}{3-q} \right) \rho_1 {R}^{-q} =  \left( \frac{3}{3-q} \right)  \rho\left({R}\right)
,\end{equation}  \\
and for a gradient typical of infall, $ q = 1.5 $, 
it gives \\

$ \langle\rho\left({R}\right)\rangle = 2 \rho_1 {R}^{-3/2}  = 2 \rho\left({R}\right.). $ 

\subsection{ionisation expansion in a decreasing power~law density profile }\label{sec:ionExpCalculation}
We consider here a decreasing density distribution with a power~law of index q and a central core of constant density to avoid singularity at the orign and mimic a core envelope 
using the same definition as in Eq.~\ref{rhoprof}:
\begin{equation}  
\label{rhoprofbis}
\rho =  \left\{  \begin{array}{rrr} 
      \rho_\text{c} & \mbox{for} & r \leq r_\text{c} \\
      \rho_\text{1} \times r^{-q} & \mbox{for} & r > r_\text{c} \\
      &\mbox{with} & \rho_\text{c} = \rho_\text{1} \times r_\text{c}^{-q}. 
   \end{array} \right .
\end{equation}
Considering the density, $\rho_i$,  and the speed, $c_i$, in the ionised medium, the speed of the shell $V$, from the Rankine-Hugoniot conditions \citep[Eq.~B.5,][]{minier2013}  we get
\begin{equation}  
\label{RH}
 \rho_i c_i^2 \approx \rho\left( r \right)  V^2
\end{equation}
using the density profile of (Eqs.~\ref{rhoprof}/\ref{rhoprofbis}), it transforms into 
\begin{equation}  
\label{RH2}
 \rho_i c_i^2 \approx \rho_\text{1} \times r^{-q}   V^2 \approx \rho_\text{c} \times \left( \frac{r}{r_c} \right)^{-q}   V^2 
.\end{equation}
Considering an initial Str\"omgren radius smaller than the core radius $r_c$, Eq.~\ref{eq_rstromgren0} and the photon conservation gives  
\begin{equation}  
\label{fluxCons}
{\rho_c}^2 {R_{\rm Str}}^{3} = {\rho_i}^2 {r}^{3} \Leftrightarrow \frac{\rho_c}{\rho_i} = \left( \frac{r}{R_{\rm Str}} \right)^{3/2} 
.\end{equation}
 Equation~\ref{RH2} can then be written as\\
\begin{equation}  
\label{Vitesse}
 V \approx \left( \frac{r_c}{r}  \right)^{-q/2} \left( \frac{R_{\rm Str}}{r}  \right)^{3/4} c_i \approx {r_c}^{-q/2} R_{\rm Str}^{3/4} c_i  \left( \frac{1}{r}  \right)^{(3-2q)/4} 
.\end{equation}
This equation can be integrated to give the radius of the shell, $r_\text{shell}$, as a function
of time with $t_c$ the time at which the shell reaches $r_c$:
\begin{equation}
\label{radiusQtime}
r_\text{shell} = r_c\left(1+\frac{7-2q}{4}\frac{R_{\rm Str}^{3/4}}{r_c^{7/4}}c_{i}\left(t-t_c\right)\right)^{4/(7-2q)}
.\end{equation}
Equation~\ref{Vitesse} taken at $r_c$ allows defining the speed at this point,
\begin{equation}
\label{Vc}
V\left( r_c \right) = V_c \approx c_i \left( \frac{R_{\rm Str}}{r_c} \right)^{3/4} 
,\end{equation}
and can be introduced in Eq.~\ref{radiusQtime}
\begin{equation}
\label{radiusQtimeBis}
r_\text{shell} = r_c\left(1+\frac{7-2q}{4}\frac{V_c}{r_c}\left(t-t_c\right) \right)^{4/(7-2q)}
.\end{equation}

\section{ Column density line of sight calculation from an envelope of decreasing density with a power~law profile }\label{sec:fromdecreasingDensity2losColDens}

\begin{table} 
\centering
 \caption{ Column density dependance to power~law index of decreasing density envelope}
\begin{tabular}{|c|c|c|c|c|}   
\hline  
\multicolumn{1}{|c|}{  }
& \multicolumn{2}{|c|}{ Integral $ I\left( a \right) $ }
& \multicolumn{2}{|c|}{} 
\\ 
q       & analytical    & numerical & $ \Sigma_1/1pc$ & $\zeta$ \\
\hline
$1$ &
$\ln \left( y + \sqrt{y ^ 2 + a ^ 2} \right) $ & (1) &    -  & -  \\ 
$3/2$ &
$2.6218/\sqrt{  a  }  $ &$2.6/\sqrt{  a  }$ & $5.2 \rho_1 $  & $5.2$  \\ 
$2$ &
$1/a \arctan \left(   y / a  \right)  $ &$ \pi / { 2  a  }  $ & $ \pi \rho_1 $  & $\pi$  \\ 
$5/2$ &
$1.1985/\sqrt{  a^3  }  $ &$1.2/\sqrt{  a^3  }$ & $2.4 \rho_1 $   & $2.4$ \\ 
$3$ &
$y/{ \left( a \sqrt{y ^ 2 + a ^ 2} \right) } $ & $1  / {   a^2  } $& $2 \rho_1 $ & $2$ \\ 
\hline  
 \end{tabular}
 \label{table_qintegral}    
\begin{list}{}{}
\item[]
{{\bf Notes:}  (1) Analytical solution diverges at $\infty$. Integration can only be done on a spatialy limited size, and analytical values will depend on the outside radius $R_{out}$.
}
\end{list}      
\vspace{-.5cm}      
\end{table}

When considering 
a decreasing density distribution with a power~law of index q : $ \rho \left({r}\right) = \rho_1 {r}^{-q} $,  
the column density as measured along the $y$ axis at impact parameter $a$  is given by
\begin{equation}  \label{cdlos1}
\Sigma(a) = 
2 \int_{0}^{\infty} \rho_1 {r}^{-q} dy =   2 \rho_1 \int_{0}^{\infty}  \left( {\sqrt{a^2 + y ^ 2}} \right) ^{-q} dy 
\end{equation}
with 
$ I\left( x \right) =    \int_{0}^{\infty}  \left( {\sqrt{x^2 + y ^ 2}} \right) ^{-q} dy=    \int_{0}^{\infty}  \left( {{x^2 + y ^ 2}} \right) ^{-q/2} dy ,$ \\
the column density at impact parameter $a$ is then given by
\begin{equation}  \label{integralelli}
\Sigma(a) =    2 \rho_1 I\left( a \right) 
.\end{equation} 
The analytical integration for integer values of q leads to well-known integrals already calculated by \cite{Yun1991}.
Half integers lead to elliptical integrals, and their tabulated values result in a polynomial expression of $I\left( a \right) $.
The analytical and numerical values are given in Cols. 2 and 3 of Table~\ref{table_qintegral}.
Adopting a value of 1 pc for the impact parameter $a$, we can derive a relation between the column density and the density via
$ \Sigma_1(a=1pc) =    2 \rho_1 I\left( a = 1 pc \right) $. 
The relation of  $\rho_1$ as a function of $ \Sigma_1$  uses the conversion factor $\zeta$ 
corresponding to the appropriate $q$ value and given in Col.5 of Table~\ref{table_qintegral}:

\begin{equation}  \label{rho1calc}
  \rho_1  =  \frac{\Sigma_1}  {  \zeta \times 1 \rm\,pc/cm } =   \frac{1000}  {  3.08  \times   \zeta }  \rm\, cm^{-3} \frac{\Sigma_1}  { 10^{21}\rm\, cm^{-2} }  
.\end{equation} 

The dependence between $\zeta$ and $q$ is given at the 5\% error level by the following relation
\begin{equation}  \label{zetaFormula}
  \zeta  =      \frac{\pi}  {  \left(  q - 1 \right)^{2/3} }  
.\end{equation}

\section{ density gradient steepening by external compression, as seen in simulations}\label{sec:griedntandcompression}
 
 \cite{Hennebelle2003} have made numerical simulations to test the influence  on density profiles of compression induced by additional external pressure.
 The compression is characterised (see their Eq.12) by a dimensionless factor ($\phi$) that gives the number of sound-crossing times needed to double the external pressure.
 High $\phi$ values correspond to low compression, called subsonic and slow pression increase.
\cite{Hennebelle2003} show radial density and velocity profile at five time steps. 
The three first ones correspond to prestellar phases, well represented by Bonnor-Ebert spheres \citep{Bonnor1956,Ebert1955}.
The fourth step would correspond to the protostellar class 0 phase and the last one to the beginning of the class I phase (see \citealt{Andre2000} for classes 0 and I definitions). 
For a supersonic compression ($\phi<$1)  during the prestellar phase, a density wave crosses the core profile towards the interior. 
For a sonic or subsonic compression ($\phi\geqslant$1), the Bonnor-Ebert sphere like profile is smoothly distorted during prestellar phases 
leading to a single power~law in protostellar phases.
 
\begin{table}  
\centering
 \caption{Envelope density power~law slope dependence on compression at two time step of protostellar stage}
\begin{tabular}{|c|c|c|c|c|c|}
\hline   
$\phi$ & 10  & 3 & 1 & .3 &  .1  
\\
\hline  
time step 4 : Class0 start & -2.1 & -2.1 & -2.2 & -1.9 & -2. 
\\
time step 5 : ClassI start & -1.5 & -1.6 & -1.65 & -1.7 & -1.8 
\\ 
\hline  
 \end{tabular}
 \label{table_phi}    
\vspace{-.1cm}
\end{table}


We measured and give in Table \ref{table_phi} the variation in the density profile power~law index (q) as a function of the compression factor $\phi$ during the two time steps 
corresponding to the beginning and the end of Class 0 phase, (steps 4 and 5). 
During the Class 0 phase, the slope of the density profile does not depend on the compression and  remains  q = -2,  
the classical value expected for hydrostatic equilibrium. 
At the end of the Class 0 phase, the slope of the power~law fitting the density profile is correlated to the compression given by the $\phi$ value (see Table~\ref{table_phi}).
It increases from 1.5, the classical value of a free-falling envelope,  to 1.8 when $\phi$  varies from 10 (virtually no compression) to .1 (strongly supersonic compression). 
This 20 \% steepening of the density profile could be one  effect that also occurs at a later stage and to larger radii as observed in the MonR2 central UC\hii region,
as advocated in Sect.\ref{sec:protinfall}.

\section{\hii region swept-up shells and  their contribution to the column density} 
\label{sec:HIIshellContribution}

The `collect-and-collapse' scenario proposes that the ionising flux of OB-type stars indeed efficiently sweeps up the gas located within the \hii region extent and develops a shell at the periphery of \hii bubbles \citep{ElmegreenLada1977}. 

\begin{figure}[htbp]
\includegraphics[height=.75\hsize]{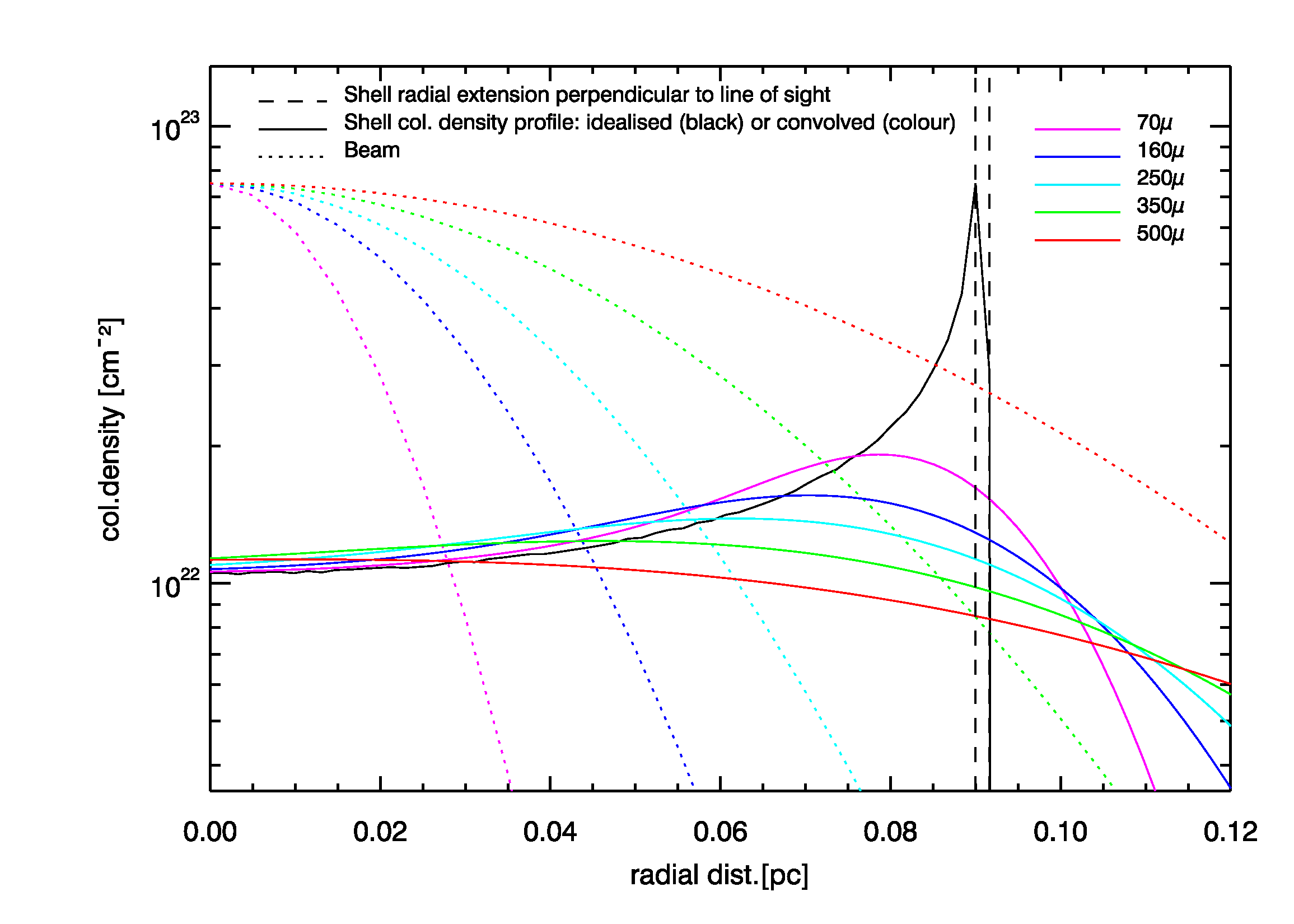}
\hfill 
\vspace{-.7cm}
\caption{Contribution of the shell to the column density profile of the central \uchii region. Its density radial extension is given by the two dashed black lines. 
The  column density resulting from the line-of-sight accumulation is illustrated by the black solid line.
The shell column density  is then convolved with the \emph{Herschel} beams (coloured dotted lines) to simulate the flux or column density of the shell (coloured solid lines) observable 
from $6\arcsec$(70\,$\mu$m) to $36\arcsec$ (500\,$\mu$m) resolutions. 
\label{fig:shellProfilCenter}}
\vspace{-.5cm}
\end{figure}

We (see also Appendix~\ref{sec:HIIshellAttenuation}) investigate the contribution of this very narrow component to the column density measured with \emph{Herschel}. 
This contribution should be especially large at the border of the \hii regions where the  line of sight tangentially crosses the shell.

We thus chose to model the density structure of a shell surrounding an \hii region, similar to the one powered by the Mon~R2-IRS1 star. 
We used a density structure that linearly increases with the radius, $\rho(r)\propto r$ (see Fig.~$~$\ref{fig:rhoProfilOnGeom})
to mimic those suggested by the calculations of \hii region expansion \citep{HosoInu2005, HosoInu2006}, 
giving more weight to the shell outer layers, and a thickness of $\sim$0.002~pc as modelled by \citet[Fig.2]{Pilleri2013}. 
\citet{pilleri2012, Pilleri2013} studied the shell of the central Mon~R2 \uchii and considered two concentric slabs with homogeneous density, located at $\sim$0.08~pc around the \hii region. 
In their model, the photo-dissociation region slab has a density of $2\times 10^5$~cm$^{-3}$ and a $6.5\times10^{-4}$~pc thickness, while the high-density shell has a $3\times10^6$~cm$^{-3}$ density and a $10^{-3}$~pc thickness. We used the densities of these slabs to model the inner and outer density values of the shell surrounding the central \uchii region.

Figure~\ref{fig:shellProfilCenter} shows the column density result of the central \uchii shell calculated with the assumptions above. 
the idealised profile of this shell (black solid line), flat towards the centre and sharply increasing to the border.  
The shell was convolved with the \herschel~beams (dotted coloured lines) to evaluate its contribution to the fluxes and column density profiles measured towards this \hii region. 
The border increase due to the shell seen tangentially is completely smeared out for all \herschel~wavelengths tracing the cold gas density, i.e. for $\lambda \ge 160\,$\microm\ (see Fig.~\ref{fig:shellProfilCenter}). 
The column density profiles measured by \herschel~have angular resolutions that are too coarse ($25\arcsec$ and $36\arcsec$), by factors of at least four. 
On top of that, the mixing along the line of sight with other density components, such as the envelope and background filaments, means that the centre-to-limb contrast of the shell is rarely observable. 
We thus cannot expect to resolve any shell around compact or ultra-compact \hii regions located at 830~pc. 
In contrast in the case of the extended northern \hii region, the relative contribution of the shell compared to the envelope is more important, increasing the centre-to-limb contrast and allowing a marginal detection of the shell (see Fig.~ \ref{fig:4compProfilNorth}). 
Therefore, the shell is not a dominant component of the column density structure of \hii regions and their surroundings, but it cannot always be neglected for its complete bubbling.

\section{Column density dilution on the central line of sight by shell}\label{sec:HIIshellAttenuation}

\subsection{attenuation by shell in homogenous envelope}\label{sec:HIIshellAttenuationHomogEnv}
We first assumed the simple scenario of a bubble expanding in an homogenous envelope where which \emph{\emph{all}} the matter initially located within a sphere with a $R_{\rm \ion{H}{ii}}$ radius concentrates in the shell located at this very same radius. 
According to numerical simulations, the \hii region's swept-up shells have small thicknesses, $L_\text{shell}<0.01$~pc \citep[e.g.][]{HosoInu2005}. 
The almost total mass transfer from the bubble to the shell thus leads to the approximate equation,
 \begin{eqnarray}  \label{eq:ShellAndBubble}
M_\text{initial} & =  &\frac {4}{3}\pi\, {R_{\rm \ion{H}{ii}}}^3 \times  \rho_\text{initial} \nonumber \\
   & = & M_\text{shell} \simeq 4 \pi \, {R_{\rm \ion{H}{ii}}}^2 \,L_\text{shell} \times \rho_\text{shell}, 
\end{eqnarray}
where $\rho_\text{initial}$ is the initial constant density of the envelope, and $\rho_\text{shell}$ the shell density, assumed to be homogeneous. 
Using the relation between $\rho_\text{shell}$ and $ \rho_\text{initial}$ given by Eq.~\ref{eq:ShellAndBubble}, 
the column density of the shell, $\Sigma_\text{shell}$, measured along the line of sight towards the centre of the region simply relates to that of the initial gas sphere, $\Sigma_\text{initial}(r<R_{\rm \ion{H}{ii}})$, with homogeneous density  $\rho_\text{initial}$ and radius $R_{\rm \ion{H}{ii}}$ through
\begin{eqnarray} \label{eq:colden}
\Sigma_\text{shell} & = & \Sigma_\text{obs}  = 2\times (L_\text{shell} \times \rho_\text{shell}) \nonumber \\
 & \simeq & 2 \times (R_{\rm \ion{H}{ii}}  \times \rho_\text{initial}\,  /3) = \Sigma_\text{initial} \,/ 3
.\end{eqnarray}
The column density measured towards the center of \hii regions with fully developed bubbles is thus expected to be divided by 
three compared to its original value: 
\begin{eqnarray} \label{eq:etaShellCstDensEnv}
\eta_\text{shell} = \frac{ \Sigma_\text{initial} }{ \Sigma_\text{obs} }  = 3. 
\end{eqnarray}
When approaching the border of the \hii region, the line of sight crosses a greater part of the shell, and the column density reaches higher values.
However, as shown in Appendix.~\ref{sec:HIIshellContribution}, the very small size expected for the shell will result in a beam dilution with very small enhancement that is usually not observable.
 
In the case of an \hii region of radius $R_{\rm \ion{H}{ii}}$ that not fully developed within an envelope of density $\rho_\text{initial}$, to reach its external radius, $R_{\rm env}$, 
the contribution of the outer residual envelope, $\Sigma_\text{env}$, needs to be accounted for. 
The initial column density is now $\Sigma_\text{initial}(r<R_{\rm env})=2\times R_{\rm env} \times \rho_\text{initial}$. 
The column density observed towards the developed \hii bubble would thus be
\begin{eqnarray}
\Sigma_\text{obs} & = & \Sigma_\text{shell}+\Sigma_\text{env} \simeq  \Sigma_\text{initial} \,/ 3\times \frac{R_{\rm \ion{H}{ii}}}{R_{\rm env}} + \Sigma_\text{initial}\times \frac{R_{\rm env} - R_{\rm \ion{H}{ii}}} {R_{\rm env}}\nonumber\\
 & \simeq & \Sigma_\text{initial} \times \frac{3\times R_{\rm env} - 2 \times R_{\rm \ion{H}{ii}}} {3\times R_{\rm env}},
\end{eqnarray}
and the decreasing factor or line-of-sight attenuation factor, $\eta$, would be 
\begin{eqnarray} \label{eq:etaCstDensEnv}
\eta = \frac{ \Sigma_\text{initial} }{ \Sigma_\text{obs} } \simeq \frac{3} {3-2 \times\frac{R_{\rm \ion{H}{ii}}}{R_{\rm env}}}.
 \end{eqnarray}
It can be noticed that for a fully extended \hii region that reaches the size of the envelope 
$R_{\rm \ion{H}{ii}} = R_{\rm env}$, the attenuation factor reaches 3, the value already obtained for the shell alone.

{These purely geometrical attenuation factors are weak for the small compact and ultra-compact \hii regions ($\eta\simeq 1.1/1.2$), 
but start to be noticeable for the more extended northern \hii region ($\eta\simeq 1.5$).}
We could estimated the density of the initial protostellar envelope before the \hii region develops, $ \rho_\text{initial}$, by correcting the mean density $\langle\rho_\text{obs}\rangle$, 
measured in Sect.~\ref{sec:HIIRenvdensprof} from observed column density through Eq.~\ref{eq:meanden}. 
The density correction due to the geometrical effect of \hii region expansion is obtained by the following relation:
$\rho_\text{initial}\sim \langle\rho_\text{obs}\rangle\times\eta$.

\subsection{attenuation in an envelope of decreasing density}\label{sec:HIIshellAttenuationDecrEnv}

In the more realistic case of an envelope with a decreasing density gradient, 
it is expected that the attenuation factor of the column density by \hii bubbles, $\eta$, should be greater than in the case of a constant-density envelope 
and constantly increasing as the \hii region expands. 

The mass collected in the shell for the expansion of the \hii region at a radius $R_{\rm \ion{H}{ii}}$ is given by Eq.~\ref{mass2}.
The density in the shell of thichness $L_\text{shell}$ is, then, 
\begin{equation}  \label{rhoShellDecEnv}
\rho_\text{shell} = \frac{M\left( R_{\rm \ion{H}{ii}} \right)}{4\pi R_{\rm \ion{H}{ii}}^{2} L_\text{shell} } =    \frac{\rho_1 R_{\rm \ion{H}{ii}}^{1-q}} {(3-q) \times  L_\text{shell} }.     \nonumber
\end{equation}  \\
The column density observed for this shell at small impact parameter, near the \hii region centre is then 
\begin{equation}  \label{coldensShellDecEnv}
 \Sigma_\text{shell} \left( R_{\rm \ion{H}{ii}} \right) ={\rho_\text{shell} } \times { L_\text{shell} } =   \rho_1  \frac{ R_{\rm \ion{H}{ii}}^{1-q} } {3-q }.     \nonumber
\end{equation}  \\
The initial value of the column density was 
\begin{equation}  \label{coldensInitShellDecEnv}
\Sigma_\text{initial} \left( R_{\rm \ion{H}{ii}} \right) =  \int_{0}^{R_{\rm \ion{H}{ii}}} \rho(r) \, dr =  \int_{0}^{R_{\rm \ion{H}{ii}}} \rho_1 {r}^{-q} dr =   \rho_1 \,  \frac {R_{\rm \ion{H}{ii}}^{1-q} } {1-q} 
.\end{equation}
This relation is valid only for $q$ < 1 when the value of ${R}^{1-q}$  near 0 is negligible.
The attenuation by the shell is given by 
\begin{eqnarray} \label{eq:etaShellDecrEnv}
\eta_\text{shell}  = \frac{ \Sigma_\text{initial} }{ \Sigma_\text{obs} } = \frac {3-q} {1-q} .
 \end{eqnarray}
 For constant density envelope ($q$=0), Eq.~\ref{eq:etaShellCstDensEnv} is recovered.  \\   
 
 To evaluate the attenuation with an existing residual outer envelope we have to calculate the corresponding column density contribution,
\begin{eqnarray} \label{coldens4resEnvInDecEnv}
 \Sigma_\text{env} \left( R_{\rm \ion{H}{ii}} \right)  = \int_{R_{\rm \ion{H}{ii}}}^{R_\text{env}} \rho(r) \, dr =  
 \frac{ \rho_1 }{1-q} R_\text{env}^{1-q}  \left( 1 - {\left( \frac {R_{\rm \ion{H}{ii}}} {R_\text{env}}\right ) }^{1-q} \right )
,\end{eqnarray}
then
\begin{eqnarray} \label{eq:etaDecrEnv-head}
\eta = \frac{ \Sigma_\text{initial}  \left( R_\text{env} \right)}{ \Sigma_\text{obs}  } 
      = \frac{ \Sigma_\text{initial}  \left( R_\text{env} \right) }{ \Sigma_\text{shell} \left( R_{\rm \ion{H}{ii}} \right) + \Sigma_\text{env} \left( R_{\rm \ion{H}{ii}} \right) } \nonumber
 \end{eqnarray}
 will give, after some calculations,
\begin{eqnarray} \label{eq:etaDecrEnv}
\eta = \frac {3-q} {3 - 2\times  {\left( \frac {R_{\rm \ion{H}{ii}}} {R_\text{env}}   \right ) }^{1-q}  - q }
.\end{eqnarray}
For a shell reaching the size of the envelope 
$R_{\rm \ion{H}{ii}} = R_{\rm env}$, Eq.~\ref{eq:etaDecrEnv} leads to Eq.~\ref{eq:etaShellDecrEnv}, which is obtained for the shell alone, and for a constant density envelope ($q$=0)
it gives Eq.~\ref{eq:etaCstDensEnv}
  
These relations are only valid for $q$ < 1, owing to the singularity at the origin of the density distribution.

\subsection{attenuation in envelope of decreasing density with a constant density  core}\label{sec:HIIshellAttenuationDecrEnvWithCore}

To avoid the singularity at the origin we need to use the density profile defined by Eq.\ref{rhoprofbis} with a central core of constant density 
 $\rho_\text{c}$ and a size of $r_\text{c}$.
Following the same calculation as above they give for 
the mass
 \begin{eqnarray}  \label{massDEWCC}
M\left( R\right) & = & \int_{0}^{R} \rho \left( r  \right) V \left( r  \right) dr \nonumber \\ 
        & = & \int_{0}^{ r_\text{c}} \rho \left( r  \right) V \left( r  \right) dr  + \int_{r_\text{c}}^{R} \rho \left( r  \right) V \left( r  \right) dr \nonumber \\
        & = & \frac{4\pi \rho_1 R^{3-q} }{ 3 - q}  \left( 1 - \frac{q}{ 3}  \left( \frac{r_\text{c}}{R}  \right)^{3-q}    \right)
,\end{eqnarray}
the density in the shell
\begin{eqnarray}  \label{rhoShellDEWCC}
\rho_\text{shell} = \frac{M\left( R_{\rm \ion{H}{ii}} \right)}{4\pi R_{\rm \ion{H}{ii}}^{2} L_\text{shell} },      \nonumber
\end{eqnarray}
the shell  column density near the centre of the \hii region
\begin{eqnarray}  \label{coldensShellDEWCC}
 \Sigma_\text{shell} \left( R_{\rm \ion{H}{ii}} \right)  & = & {\rho_\text{shell} } \times { L_\text{shell} } =   \frac{M\left( R_{\rm \ion{H}{ii}} \right)}{4\pi R_{\rm \ion{H}{ii}}^{2} }   \nonumber  \\
        & = &   \frac{  \rho_1 R^{1-q} }{ 3 - q}  \left( 1 - \frac{q}{ 3}  \left( \frac{r_\text{c}}{R_{\rm \ion{H}{ii}}}  \right)^{3-q}    \right)
,\end{eqnarray}  \\
the initial value of the column density 
\begin{equation}  \label{coldensInitShellDEWCC}
\Sigma_\text{initial} \left( R_{\rm \ion{H}{ii}} \right) =  \frac{  \rho_1 R_{\rm \ion{H}{ii}}^{1-q} }{ 1 - q}  \left( 1 - q \left( \frac{r_\text{c}}{R_{\rm \ion{H}{ii}}}  \right)^{1-q}    \right)
,\end{equation}
and then the attenuation by the shell  
\begin{eqnarray} \label{eq:etaShellDEWCC}
\eta_\text{shell}  = \frac{ \Sigma_\text{initial} }{ \Sigma_\text{obs} }  
= \frac {3-q} {1-q}  \times
            \frac { 1 - q                    \left(  {r_\text{c}} / {R_{\rm \ion{H}{ii}}}  \right)^{1-q}    }
                    { 1  -  \frac{q}{ 3}   { \left(  {r_\text{c}} /{R_{\rm \ion{H}{ii}}}  \right)}^{3-q} } 
.\end{eqnarray}
All these relations reduces to the equivalent one of the previous section for $r_\text{c}$=0.

The same calculations as before for an existing residual outer envelope are still applicable here. 
As shown by \citet{Franco1990}, the shell develops only 
for $q$< 1.5, so  $q$< 3, and the terms $\left( \frac{r_\text{c}}{R_{\rm \ion{H}{ii}}}  \right)^{3-q}$ in the equations above are always negligible.
Then the calculations give
\begin{eqnarray} \label{eq:etaDEWCC}
\eta = \frac {3-q} {3 - 2\times  {\left( \frac {R_{\rm \ion{H}{ii}}} {R_\text{env}}   \right ) }^{1-q}  - q } \times \left( 1 - q \left( \frac{r_\text{c}}{R_{\rm \ion{H}{ii}}}  \right)^{1-q}    \right)
\end{eqnarray}
 for the attenuation, and for $r_\text{c}$=0 we recover Eq.~\ref{eq:etaDecrEnv}

The application to the northern \hii region with appropriate 
values of the different parameters gives an attenuation factor $\eta \simeq 25$,
which becomes $\sim$ 100 when the \hii region expansion reaches the size of the envelope (${R_{\rm \ion{H}{ii}}} = {R_\text{env}}$).

The formal treatment of the case of the compact and UC \hii regions would require considering an envelope with two density gradients, 
but the attenuation would be as efficient as for the northern extended \hii region once expansion of these region occurred.

\end{appendix}

\clearpage

\end{document}